\documentstyle[12pt]{article}
\newcommand{\dd}{{\rm d\kern-.17em I}}
\newcommand{\w}{\wedge \kern-.8em \wedge}

\newcommand{\real}{{\rm I\kern-.17em R}}

\newcommand{\beu}{
\begin{picture}(11,3)(-2,-2)
\put(0,8){$\scriptscriptstyle{(1)}$} \put(0,-2){$\beta$}
\end{picture}}

\newcommand{\alu}{
\begin{picture}(11,3)(-2,-2)
\put(0,6){$\scriptscriptstyle{(1)}$} \put(0,-2){$\alpha$}
\end{picture}}

\newcommand{\ou}{
\begin{picture}(11,3)(-2,-2)
\put(0,6){$\scriptscriptstyle{(1)}$} \put(0,-2){$\omega$}
\end{picture}}

\newcommand{\od}{
\begin{picture}(11,3)(-2,-2)
\put(0,6){$\scriptscriptstyle{(2)}$} \put(0,-2){$\omega$}
\end{picture}}

\newcommand{\ad}{
\begin{picture}(11,3)(-2,-2)
\put(0,6){$\scriptscriptstyle{(2)}$} \put(0,-2){$a$}
\end{picture}}

\newcommand{\bd}{
\begin{picture}(11,3)(-2,-2)
\put(0,8){$\scriptscriptstyle{(2)}$} \put(0,-2){$b$}
\end{picture}}

\begin{document}

\baselineskip=22pt plus 0.2pt minus 0.2pt \lineskip=22pt plus
0.2pt minus 0.2pt
\begin{center}
\vspace*{1cm} \LARGE
Quadratic $s$-Form Field Actions with Semi-bounded Energy.\\

\vspace*{1.5cm}

\large

J.\ Fernando\ Barbero\ G.\footnote{barbero@laeff.esa.es},\\and\\
Eduardo J. S. Villase\~nor\footnote{ejesus.sanchez@mat.ind.uem.es}

\vspace*{1.5cm}

\normalsize
{\it Escuela Superior de Ingenier\'{\i}a Industrial,\\
Universidad Europea \\
Urb. El Bosque, C/ Tajo s/n\\
Villaviciosa de Od\'on, Madrid, 28670\\
Spain\\\hspace{5mm}\\}

\vspace{.3in}
July 20, 2000\\
\vspace{.3in} ABSTRACT
\end{center}

We give in this paper a partial classification of the
consistent quadratic gauge actions that can be written in terms
of $s$-form fields. This  provides a starting point to study the 
uniqueness of the Yang-Mills action as a deformation of Maxwell-like theories. We also show that it is impossible to write kinetic 1-form terms
that can be consistently added to other 1-form actions such as tetrad gravity in four space-time dimensions even in the presence of a 
Minkowskian metric background.

\vspace*{1cm} \noindent PACS number(s): 11.15-q, 03.65.Ca, 04.20.Fy 


\pagebreak

\setcounter{page}{1}

\section{Introduction}

During the last years a great deal of effort has been devoted to
the problem of describing consistent quantum field theories
obtained by deformations of well known free Lagrangians such as
the Maxwell action for electromagnetism and generalizations to $p$-forms \cite{YM}-\cite{HennK}, the
Fierz-Pauli \cite{FP} model for spin 2 fields in a Minkowskian
metric background and many others. In all these cases the starting point 
is the same: take several copies of a free action such as the ones quoted
above and study the possible interaction terms that can be consistently
added to them. Consistency in this context means that the number
of physical degrees of freedom described by the free and deformed
actions are the same and the gauge symmetries and the algebra of
gauge transformations reduce to the free ones when taking the
coupling constants to zero. The main results described in the
previous papers are the uniqueness of the Yang-Mills action and
the fact that consistent gravitational theories involving several
metrics reduce to the addition of non-interacting copies of
general relativity. In our opinion, there is a point that needs to
be clarified: to which extent the free actions taken there as starting points are the most general ones. It is a somewhat surprising fact to realize that a complete characterization of free actions is not available. The purpose of this paper is to study a rather general class of them 
as a step towards its complete classification\footnote{Some results in
this direction, concerning quadratic diff-invariant Lagrangians, have
already been published \cite{Nos}.}.

In the previous examples the free actions considered are
physically consistent because they have a semi-bounded energy and,
hence, a well defined vacuum. This feature is kept after the
introduction of interaction terms, at least for small values of the
coupling constant. This leads us to demand, as a first
requirement, that the free actions that we study here must satisfy that
the energy be semi-bounded\footnote{An unbounded energy at the
quadratic level is found in the Higgs model due to the negative
mass parameter, however  this is compensated by the higher degree
terms in the quartic Higgs potential. We cannot rely on this kind
of mechanism in the kinetic case because we do not want to have
terms involving more than two derivatives.}. A first choice that
we must make is the type of fields that we want to work with. As
shown in \cite{Nos} if one demands diff-invariance of a quadratic
action one is forced to consider $s$-form fields and the only
available derivative operator is the exterior differential. One
can give up diff-invariance by introducing a metric background. In
this case the adjoint exterior differential (built with the help
of the Hodge dual defined by the background metric) is also
available. Of course, other types of geometrical objects can be
considered --general tensor fields of arbitrary valence-- but, to
stay within the framework of Yang-Mills theories, this paper will
deal only with 1-forms (although we will consider generalizations
to $s$-form fields). Notice, however, that even in this restrictive
setting some questions may be posed --and answered-- concerning
other types of theories, such as gravity, that can also be written
purely in terms of 1-forms.

\bigskip

According to the discussion above we consider in this paper
the most general quadratic local action that can be written in
terms of 1-forms, the exterior differential and its dual in a
Minkowskian space-time. We will avoid the use of mass parameters
as they are expected to lead to ill behaved propagators for large
momenta. As the type of analysis presented in the paper can be
extended to $s$-forms without extra effort we will consider in
section II an action depending on an arbitrary number of $s$-form
fields with two constant arbitrary matrices $P$ and $Q$ that
partially generalizes the action for 1-forms. In order to extract
its physical content and describe its gauge symmetries one may
rely on the standard  Hamiltonian analysis. This, however, proves
to be a  cumbersome way to attack the problem because the
secondary constraints that appear at several stages depend on the
algebraic properties of $P$ and $Q$ in a non-trivial way. A
superior strategy, as shown in \cite{Nos}, is that of using
covariant symplectic techniques \cite{CR1}-\cite{Zuc}. They are
based on the direct study of the space of solutions to the field
equations and the symplectic structure defined in it. The fact
that we are dealing with quadratic actions (and, hence, {\it
linear} field equations), will allow us to solve them completely
and parametrize the solutions in a very convenient way. With these
solutions in hand it is possible to obtain the symplectic
structure that provides us with  both a concrete description of
the reduced phase space and the gauge symmetries present in the
model. This is discussed in section III.

Once we know what the physical modes are, we want to characterize
them and select the matrices $P$ and $Q$ in order  to have a
consistent theory in the sense described before. To this end we
obtain in section IV, also by using symplectic techniques, the
energy-momentum and angular momentum. The main result of the paper
is the partial classification of the consistent quadratic
lagrangians in four dimensions that can be written in terms of
$s$-forms. This classification is partial because some additional
terms can be added for 2-form fields and we do not consider free
actions with cross-terms involving different types of
forms. Despite of this, the result may be useful as the
starting point to study their deformations along the lines
presented in \cite{Henn} and set the uniqueness of Yang-Mills on a
firmer footing. Another result of the paper is that it is
impossible to find kinetic terms that can be consistently added to
other 1-form actions such as the tetrad action for general
relativity\footnote{Here $e^I$ is a $SO(1,3)$ valued 1-form and $F_{IJ}$ 
is the curvature of a $SO(1,3)$ spin connection $A_I^{\;\;J}$.}
\begin{eqnarray}
\int e^I\wedge e^J\wedge F_{IJ}(A) \quad, \label{001}
\end{eqnarray}
even in the presence of a background metric such as Minkowski.
This would have provided a novel way to study general relativity
as an interaction term of a consistent free theory. We discuss 
this and give additional comments and conclusions in section V 
and leave computational details for the appendices.

\section{The Action, Field Equations, and Solutions.}

We start by considering the most general quadratic, second order
action that can be written in terms of 1-forms in four space-time
dimensions, without the introduction of mass terms, and using the
exterior differential $d$ and its dual $\delta$.
\begin{eqnarray}
S_1[A]=\int_{\cal M}\left[dA^\texttt{t} \wedge *PdA+\delta A^\texttt{t}
\wedge *Q\delta A \right]\quad.\label{002}
\end{eqnarray}
    Here $\cal M$ is a four dimensional pseudo-Riemannian
manifold\footnote{In the following we are
going to work with $g_{ab}=\eta_{ab}={\rm diag}\,(-,+,+,+)$, the
Minkowski metric in four dimensions, so we will chose ${\cal
M}=\real^4$. We will denote $\real^4$ indices as $a,b,\ldots$
spatial indices as $i,j,\ldots$ and $0$ will be the time index.}
without boundary with metric $g$ that defines the Hodge dual $*$,
$A$ is a set of $N$ 1-form fields that we write as a column vector
whose transpose will be denoted by $A^\texttt{t}$; $d$ is the exterior
derivative, $\delta$  its dual and $\wedge$ the usual exterior
product (we provide a dictionary to translate between form
notation and tensor notation in Appendix A).  $P$ and $Q$ are
quadratic forms represented by symmetric, real, $N\times N$ matrices.
Notice that we cannot write a quadratic second order action with
1-forms only without the use
of a background metric because all possible terms would be total
derivatives and would cancel if $\cal M$ has no boundary. As is
well known from the study  of normal modes in coupled harmonic
oscillators if $P$ or $Q$ are positive definite they can be
simultaneously diagonalized; in fact, one can find a non-singular
linear  redefinition of the fields that takes one of them to the
identity matrix and diagonalizes the other. In this case $S_1[A]$
reduces to a sum of Maxwell actions with $(\partial_aA^a)^2$
terms that is ill defined both from the classical and the quantum
point of view. If, however, $P$ and $Q$ are non-definite (for
example singular) it may be impossible to simultaneously
diagonalize  them (as can be seen in simple examples), hence, one
could expect that qualitatively new behaviors may occur in this
case. According to this, we will allow ${\rm Ker}\, P$ or ${\rm
Ker}\, Q$ to be different from $\{0\}$. In general we can have
${\rm Ker}\, P \cap {\rm Ker}\, Q\neq \{0\}$ but then we can
eliminate a set of fields from (\ref{002}) by a linear
non-singular field redefinition. To avoid this trivial situation
we demand that ${\rm Ker}\, P \cap {\rm Ker}\, Q= \{0\}$.

    The formalism that we will use in the following is powerful enough to
allow us to study generalizations of (\ref{002}) for $s$-forms
fields, so we will consider a slight modification\footnote{
Notice, however, that if we mix different values of $s$ or if
$s=2$ it is possible to add other types of terms to this action.
As our primary goal is the study of the 1-form case we will not
consider them here.} of (\ref{002})
\begin{eqnarray}
S_s[A]=\int_{\real^4}\left[dA^\texttt{t} \wedge *PdA+ \delta
A^\texttt{t}\wedge *Q\delta A\right]\quad\label{003}
\end{eqnarray}
where now $A$ is a set of $N$ $s$-form fields. The field
equations obtained from (\ref{003}) by performing variations with respect
to $A$ are
\begin{eqnarray}
P\delta d A+Qd\delta A=0\quad.\label{004}
\end{eqnarray}
This is a system of second order partial differential equations.
The first step to solve them consists of introducing linear bases
for ${\rm Ker}\, P$, ${\rm Ker}\, Q$ and complete them to obtain
a basis for $\real^N$. This will allow us to get a convenient set
of equations from (\ref{004}) that can be separately studied. We
choose sets of linearly independent vectors\footnote{The letters
$p,q,r$ will be used as indices within the different subspaces
defined by $\{e_p\}$, $\{e_q\}$, and $\{e_r\}$.} $\{e_p\}$,
$\{e_q\}$, and $\{e_r\}$ such that ${\rm Ker}\,P={\rm
Span}\,\{e_p\}$, ${\rm Ker}\, Q={\rm Span}\,\{e_q\}$,  and
$\real^N={\rm Span}\, \{e_p,e_q,e_r\}$ and write $A=A^p e_p+A^q
e_q+A^r e_r$. Notice that the condition ${\rm Ker}\, P\cap {\rm
Ker}\, Q=\{0\}$ implies that $\{e_p,e_q\}$ are linearly
independent. We can rewrite (\ref{004}) as
\begin{eqnarray}
P\delta d(A^qe_q+A^re_r)+Qd\delta (A^pe_p+A^re_r)=0\quad.\label{005}
\end{eqnarray}
A set of necessary conditions that the solutions to (\ref{005})
must satisfy can be found by acting on it with either $d$ or
$\delta $ and using $d^2=0$, $\delta^2=0$. This way we get
\begin{eqnarray}
Pd\delta d(A^qe_q+A^re_r)&=&0\label{006}\\
Q\delta d \delta (A^pe_p+A^re_r)&=&0\quad.\nonumber
\end{eqnarray}
As $\{Pe_q,Pe_r\}$ are linearly independent
vectors\footnote{$\lambda^q P(e_q)+\mu^r P(e_r)=0\Rightarrow
\lambda^q e_q+\mu^r e_r\in {\rm Ker}\, P\Rightarrow \lambda^q
e_q+\mu^r e_r+\sigma^p e_p=0\Rightarrow
\lambda^q=\mu^r=\sigma^p=0.$} (and also $\{Qe_p,Qe_r\}$) we can
write (\ref{006}) as
\begin{eqnarray}
d\delta dA^q&=&\Box dA^q=0\label{007}\\
\delta d\delta A^p&=&\Box \delta A^p=0\nonumber\\
d\delta dA^r&=&\Box dA^r=0\nonumber\\
\delta d\delta A^r&=&\Box \delta A^r=0\nonumber
\end{eqnarray}
where $\Box \equiv d\delta +\delta
d=\partial_0^2-\vec{\partial}^2$ is the wave operator (see
Appendix A) that should not be confused with the Laplace operator
--notice that we work  with a Minkowskian metric--. This fact
prevents us from using the Hodge decomposition for the $s$-forms
in (\ref{007}); however a suitable decomposition, that helps in
solving these equations, may be found as follows. Consider
\begin{eqnarray}
\delta (d\alpha-A)=0\label{008}
\end{eqnarray}
for a given $s$-form $A$ and an unknown $(s-1)$-form $\alpha$.
Making the ansatz $ \alpha=\delta \theta$  (equivalent in
$\real^4$ to $\delta \alpha=0$) the previous equation writes
$\delta (\Box \theta -A)=0$ so by choosing $\theta$ as a solution
to the inhomogeneous wave equation $\Box\theta=A$ (that can
always be solved under reasonable regularity conditions) we can
find {\it some} $\alpha$ satisfying (\ref{008}). As (\ref{008})
implies the existence of a $(s+1)$-form $\beta$ such that
$d\alpha-A=-\delta\beta$ we see that given $A$ it is always
possible to find forms $\alpha$ and $\beta$ such that
\begin{eqnarray}
A=d\alpha+\delta \beta\quad.\label{009}
\end{eqnarray}
Now we are ready to solve the equations appearing  in (\ref{007}).

\subsection*{Solutions to $\Box d A^q=0$ for a $s$-form $A^q$.}

Introducing (\ref{009}) in $\Box dA^q=0$ we find $\Box d\delta
\beta^q=0\Leftrightarrow d\Box \delta \beta^q=0$. In $\real^4$
the last equation implies $\Box \delta \beta^q=d\sigma^q$ for
some $\sigma^q$. This is an inhomogeneous wave equation for
$\delta \beta^q$ that gives $\delta \beta^q=\gamma^q+d\tau^q$ with
$\Box \gamma^q=0$ and $\Box \tau^q=\sigma^q$, so finally
\begin{eqnarray}
A^q=\gamma^q+d(\alpha^q+\tau^q)\equiv
\gamma^q+d\Lambda^q\label{011}
\end{eqnarray}
where $\Lambda^q$ can be taken to be arbitrary as $\alpha^q$
itself may be arbitrarily chosen because $A^q$ enters the
equation $\Box dA^q=0$ through the combination $dA^q$.

\subsection*{Solutions to $\Box \delta A^p=0$ for a $s$-form
$A^p$.}

Introducing (\ref{009}) in $\Box \delta A^p=0$ we find $\Box
\delta d\alpha^p=0\Leftrightarrow \delta \Box d\alpha^p=0$. In
$\real^4$ this last equation implies $\Box d\alpha^p=\delta
\sigma^p$ for some $\sigma^p$. This inhomogeneous wave equation
for $d\alpha^p$ gives $d\alpha^p=\gamma^p+\delta \tau^p$ with
$\Box \gamma^p=0$ and $\Box \tau^p=\sigma^p$, so we conclude
\begin{eqnarray}
A^p=\gamma^p+\delta (\tau^p+\beta^p)\equiv \gamma^p+\delta
\Theta^p\label{013}
\end{eqnarray}
where, as before, we can take $\Theta^p$ arbitrary because
$\beta^p$ can be chosen to be arbitrary too, as suggested by the
equation $\Box \delta A^p=0$.

\subsection*{Solutions to $\Box dA^r=0$ and $\Box \delta
A^r=0$.}

$A^r$ satisfies the equation discussed in the first place so we
can always parametrize it as $A^r=\gamma^r+d\mu^r$ for some
arbitrary $\mu^r$--at this stage-- and $\gamma^r$ satisfying $\Box
\gamma^r=0$. Plugging this into the second equation $\Box \delta
A^r=0$ gives the following fourth order equation for $\mu^r$
\begin{eqnarray}
\Box\delta d\mu^r=0\quad.\label{014}
\end{eqnarray}
By using the decomposition (\ref{009}) for $\mu^r$ we write it as
$\mu^r=\delta \varepsilon^r+d\theta^r$  and, hence, (\ref{014})
gives $\Box \delta d\delta \varepsilon^r=0\Rightarrow \delta
\Box^2\varepsilon^r=0$ whose solutions have the form
$\varepsilon^r=\gamma_{{\scriptscriptstyle H\!D}}^r+\delta\beta^r$
(with $\Box^2\gamma_{{\scriptscriptstyle H\!D}}^r=0$). We
conclude that $\mu^r=\delta \gamma_{{\scriptscriptstyle
H\!D}}^r+d\theta^r$ and then
\begin{eqnarray}
A^r= \gamma^r+d\delta \gamma_{{\scriptscriptstyle
H\!D}}^r\quad.\label{017}
\end{eqnarray}
Notice that we have no arbitrariness in $A^r$.

\bigskip

    As (\ref{007}) are only necessary conditions we know that every
solution to the field equations (\ref{004},\ref{005}) can be
parametrized with the help of (\ref{011}), (\ref{013}), and
(\ref{017}) but (\ref{005}) imposes some further restrictions on
$\gamma^q$, $\Lambda^q$, $\gamma^p$, $\Theta^p$, $\gamma^r$,
$\gamma_{{\scriptscriptstyle H\!D}}^r$ given by
\begin{eqnarray}
P\delta d(\gamma^qe_q+\gamma^re_r)+Qd\delta(\gamma^p e_p
+\gamma^re_r+d\delta \gamma_{{\scriptscriptstyle H\!D}}^r
e_r)=0\quad.\label{018}
\end{eqnarray}
Though this last equation may look as complicated as the original
field equations it is, in a sense that we make precise below, a
simple {\it algebraic} equation that can be easily handled. Anyway
some simplifications are already evident because the arbitrary
objects $\Lambda^q$ and $\Theta^p$ do not appear in it. In order
to proceed further we need to parametrize $\gamma^q$, $\gamma^p$,
$\gamma^r$, and $\gamma_{{\scriptscriptstyle H\!D}}^r$. To this
end we take an inertial coordinate system $(\vec{x},t)$ in
$\real^4$ and define spatial Fourier transforms as
\begin{eqnarray}
f(\vec{x},t)=\frac{1}{(2\pi)^{3/2}}\int_{\real^3}\frac{d^3\vec{k}}{w}
f(\vec{k},t)e^{i\vec{k}\cdot \vec{x}} \quad;\quad
\frac{1}{w}f(\vec{k},t)=\frac{1}{(2\pi)^{3/2}}\int_{\real^3}d^3\vec{x}
f(\vec{x},t)e^{-i\vec{k}\cdot \vec{x}}\label{019}
\end{eqnarray}
with\footnote{We use 3-dimensional vector notation in $\real^3$
and denote the usual Euclidean scalar product with a dot.}
$w=+\sqrt{\vec{k}\cdot \vec{k}}$ introduced in the definition
(\ref{019}) in order to have an explicit Lorentz covariant
measure. We purposely use the same letter to represent a field
and its Fourier transform. We will distinguish them by their
arguments. As we will be dealing with real fields we must have
$f(\vec{k},t)=\bar{f}(-\vec{k},t)$ where in the following the bar
denotes complex conjugation. We also need to perform a $3+1$
decomposition of the various $s$-forms to obtain the time and
space components.  For a $s$-form $\omega$ with components
$\omega_{a_1\cdots a_s}$  we only need to consider
$\omega_{i_1\cdots i_s}$ and $\omega_{0i_1\cdots i_{s-1}}$ with
Fourier transforms given by
\begin{eqnarray}
\omega_{i_1\cdots i_s}(\vec{k},t)
    &=&ik_{[i_1}\alpha_{i_2\cdots i_s]}(\vec{k},t)+\beta_{i_1\cdots i_s}(\vec{k},t)\label{020}\\
\omega_{0i_1\cdots i_{s-1}}(\vec{k},t)
    &=&ik_{[i_1}a_{i_2\cdots i_{s-1}]}(\vec{k},t)+b_{i_1\cdots i_{s-1}}(\vec{k},t)\nonumber
\end{eqnarray}
and $\alpha_{i_1\cdots i_{s-1}}(\vec{k},t)$, $\beta_{i_1\cdots
i_s}(\vec{k},t)$, $a_{i_1\cdots i_{s-2}}(\vec{k},t)$,
$b_{i_1\cdots i_{s-1}}(\vec{k},t)$ satisfying the transversality
conditions $k_{i_1}\alpha_{i_1\cdots i_{s-1}}(\vec{k},t)=0$,... In
the following we will give the components of a certain form (such
as the previous one) by enclosing them between curly brackets as
follows
\begin{eqnarray*}
\omega=\left\{\begin{array}{l}ik_{[i_1}\alpha_{i_2\cdots
i_s]}(\vec{k},t) +\beta_{i_1\cdots i_s}(\vec{k},t)\\ \\
ik_{[i_1}a_{i_2\cdots i_{s-1}]}(\vec{k},t)+b_{i_1\cdots
i_{s-1}}(\vec{k},t)
\end{array}\right\}\quad.
\end{eqnarray*}
Solutions to the wave equation $\Box \gamma=0$ can be
parametrized  as in (\ref{020}) with
\begin{eqnarray*}
w\alpha_{i_1\cdots i_{s-1}}(\vec{k},t)&=&\alpha_{i_1\cdots
i_{s-1}}(\vec{k})e^{-iwt}+
\bar{\alpha}_{i_1\cdots i_{s-1}}(-\vec{k})e^{iwt}\\
\beta_{i_1\cdots i_s}(\vec{k},t) &=&\beta_{i_1\cdots i_s}(\vec{k})e^{-iwt}+
\bar{\beta}_{i_1\cdots i_s}(-\vec{k})e^{iwt}\\
wa_{i_1\cdots i_{s-2}}(\vec{k},t)&=&a_{i_1\cdots
i_{s-2}}(\vec{k})e^{-iwt}+
\bar{a}_{i_1\cdots i_{s-2}}(-\vec{k})we^{iwt}\\
b_{i_1\cdots i_{s-1}}(\vec{k},t)&=&b_{i_1\cdots
i_{s-1}}(\vec{k})e^{-iwt}+\bar{b}_{i_1\cdots
i_{s-1}}(-\vec{k})e^{iwt}\quad,
\end{eqnarray*}
where $\alpha_{i_1\cdots i_{s-1}}(\vec{k})$, $\beta_{i_1\cdots
i_s}(\vec{k})$, $a_{i_1\cdots i_{s-2}}(\vec{k})$, $b_{i_1\cdots
i_{s-1}}(\vec{k})$ are arbitrary transversal functions of
$\vec{k}$ only. If we consider now the solutions for $A^q$ given
by (\ref{011}) we see that the arbitrariness in $\Lambda^q$
allows us to absorb a piece of $\gamma^q$ in $\Lambda^q$
(precisely that piece of  $\gamma^q$ that can be written as the
exterior differential of something), and hence we can write
$\gamma^q$ as
\begin{eqnarray}
\gamma^q=\left\{\begin{array}{c}\beta^q_{i_1\cdots
i_s}(\vec{k})e^{-iwt}+\bar{\beta}^q_{i_1\cdots
i_s}(-\vec{k})e^{iwt}\\ \\b^q_{i_1\cdots
i_{s-1}}(\vec{k})e^{-iwt}+\bar{b}^q_{i_1\cdots
i_{s-1}}(-\vec{k})e^{iwt}\end{array}\right\}\quad.\label{023}
\end{eqnarray}
For $A^p$ we can absorb the piece of $\gamma^p$  that can be
written as the adjoint exterior derivative of something in the
arbitrary $\delta \Theta^p$ term and thus
\begin{eqnarray*}
\gamma^p=\left\{\begin{array}{c}\frac{i}{w}k_{[i_1}\left[\alpha^p_{i_1\cdots
i_{s-1}]}(\vec{k})e^{-iwt}+\bar{\alpha}^p_{i_1\cdots
i_{s-1}]}(-\vec{k})e^{iwt} \right]\\
\\\frac{i}{w}k_{[i_1}\left[a^p_{i_1\cdots
i_{s-2}]}(\vec{k})e^{-iwt}+\bar{a}^p_{i_1\cdots
i_{s-2}]}(-\vec{k})e^{iwt}\right]
\end{array}\right\}\quad.
\end{eqnarray*}
In the case of $A^r$ there are no arbitrary functions, however,
there is still some freedom to ``move'' pieces of $\gamma^r$ to
$\gamma^r_{{\scriptscriptstyle H\!D}}$ because every solution to
$\Box \gamma^r=0$ satisfies $\Box^2\gamma^r=0$ and
$\gamma^r_{{\scriptscriptstyle H\!D}}$ appears through $d\delta
\gamma^r_{{\scriptscriptstyle H\!D}}$. We can, in fact write
\begin{eqnarray}
\gamma^r=\left\{\begin{array}{c}\beta^r_{i_1\cdots
i_s}(\vec{k})e^{-iwt}+
\bar{\beta}^r_{i_1\cdots i_s}(-\vec{k})e^{iwt}\\\\
\frac{i}{w}k_{[i_1}\left[a^r_{i_1\cdots
i_{s-2}]}(\vec{k})e^{-iwt}+\bar{a}^r_{i_1\cdots i_{s-2}]}
(-\vec{k})e^{iwt}\right]
\end{array}\right\}\label{025}
\end{eqnarray}
and
\begin{eqnarray}
\gamma^r_{{\scriptscriptstyle H\!D}}= \left\{\begin{array}{l}
\frac{i}{w}k_{[i_1} \left[\alpha^r_{i_2\cdots
i_s]}(\vec{k})+wt\sigma^r_{i_2\cdots
i_s]}(\vec{k})\right]e^{-iwt}+\\
\hfill + \frac{i}{w}k_{[i_1}\left[\bar{\alpha}^r_{i_2\cdots
i_s]}(-\vec{k})+wt\bar{\sigma}^r_{i_2\cdots
i_s]}(-\vec{k})\right]e^{iwt}
\\ \\
\frac{1}{s}\left[ \sigma^r_{i_1\cdots
i_{s-1}}(\vec{k})+i\alpha^r_{i_1\cdots
i_{s-1}}(\vec{k})+iwt\sigma^r_{i_1\cdots
i_{s-1}}(\vec{k})\right]e^{-iwt} +\\
\hfill+ \frac{1}{s}\left[ \bar{\sigma}^r_{i_1\cdots
i_{s-1}}(-\vec{k})-i\bar{\alpha}^r_{i_1\cdots
i_{s-1}}(-\vec{k})-iwt\bar{\sigma}^r_{i_1\cdots
i_{s-1}}(-\vec{k})\right]e^{iwt}
\end{array}\right\}\quad.\label{025bis}
\end{eqnarray}
The detailed derivation of (\ref{025}) and (\ref{025bis}) is given in Appendix B.

\noindent We can now plug (\ref{023})-(\ref{025bis}) into (\ref{018}) to get the
final solution to the field equations. By doing this we get the conditions
\begin{eqnarray*}
& & P\left[isw\left( b^q_{i_1\cdots i_{s-1}}(\vec{k})e_q+
b^r_{i_1\cdots i_{s-1}}(\vec{k})e_r \right)\right]+\nonumber\\
& &\hfill+ Q\left[ w\alpha^p_{i_1\cdots i_{s-1}}(\vec{k})e_p - is
w \left(b^r_{i_1\cdots i_{s-1}}(\vec{k})
+2 \sigma^r_{i_1\cdots i_{s-1}}(\vec{k})\right) e_r\right]=0\, ,\nonumber\\
& & \nonumber\\
& & P\left[w^2\left( b^q_{i_1\cdots i_{s-1}}(\vec{k})e_q+
b^r_{i_1\cdots i_{s-1}}(\vec{k})e_r \right)\right]+\nonumber\\
& &\hfill+ Q\left[ -iw^2 s^{-1}\alpha^p_{i_1\cdots
i_{s-1}}(\vec{k})e_p -w^2 \left(b^r_{i_1\cdots i_{s-1}}(\vec{k})
+2\sigma^r_{i_1\cdots i_{s-1}}(\vec{k})\right)
e_r\right]=0\nonumber\, ,
\end{eqnarray*}
that give the algebraic equation
\begin{eqnarray}
& & P\left[
b^q_{i_1\cdots i_{s-1}}(\vec{k})e_q+
b^r_{i_1\cdots i_{s-1}}(\vec{k})e_r
\right]=\label{027}\\
& & \hspace{3cm} = Q\left[ \frac{i}{s}\alpha^p_{i_1\cdots
i_{s-1}}(\vec{k})e_p +b^r_{i_1\cdots i_{s-1}}(\vec{k})e_r
+\frac{2}{s}\sigma^r_{i_1\cdots i_{s-1}}(\vec{k})e_r \right]\,
.\nonumber
\end{eqnarray}
This equation constraints the possible values of $b^q_{i_1\cdots
i_{s-1}}(\vec{k})$, $b^r_{i_1\cdots i_{s-1}}(\vec{k})$,
$\alpha^p_{i_1\cdots i_{s-1}}(\vec{k})$, $\sigma^r_{i_1\cdots
i_{s-1}}(\vec{k})$; only a subset of these can be chosen as
independent objects. The details of this computation can be found
in Appendix C.  Once we have a complete parametrization of
the solutions to the field equations we must study their physical
content; in particular we want to know which of the arbitrary
functions describing the solution label physical degrees of
freedom and which are only gauge parameters. To this end we must
obtain the symplectic structure in the space of solutions.

\section{The Symplectic Structure: Gauge Transformations and
Physical Degrees of Freedom.}

We have obtained in the previous section the general solution to
the field equations  (\ref{004}). This solution depends on a set
of arbitrary, time dependent functions $\Lambda^q(\vec{k},t)$
and  $\Theta^p(\vec{k},t)$, and on a set of arbitrary functions
of $\vec{k}$ satisfying simple algebraic constraints. This
section is devoted to the computation of the symplectic structure
$\Omega_{{\cal S}}$ in the space of solutions to the field
equations ${\cal S}$. $\Omega_{{\cal S}}$ will provide us with
several important pieces of information:

i) It will allow  us to identify the physical degrees of freedom
in the model and the gauge transformations\footnote{It may be
argued that it is possible to identify the gauge parameters
directly from the solutions to the field equations due to their
arbitrary time dependence; it is less obvious that the remaining
$\vec{k}$-dependent functions are in one to one correspondence
with physical degrees of freedom.}.

ii) It will allow us to define the Poisson brackets in the
reduced phase space; i.e. we will  identify canonically
conjugated pairs of variables. This is a necessary  step towards
the quatization of the model.

iii) The symplectic structure can be used in a very effective way
to obtain conserved quantities such as the energy-momentum and
angular momentum that we will use in order to impose consistency
requirements on the family of models given by (\ref{003}).

    The action (\ref{003}) defines a symplectic two-form $\Omega_{{\cal F}}$ in the
space of fields ${\cal F}$ coordinatized by $A(x)$
\begin{eqnarray}
\Omega_{{\cal F}}=\int_{\real^3}J=\int_{\real^3}\left[\dd
A^{\texttt{t}} \w *Pd\dd A + (Q\delta \dd A)^\texttt{t}\w*\dd A
\right]\quad.\label{028}
\end{eqnarray}
We must distinguish between the ordinary exterior differential in
$\real^4$ ($d$) and the exterior differential in $\cal F$ ($\dd$).
In the same way we must make a distinction between the wedge
product in both cases ($\wedge$ and ${\wedge \kern-.57em \wedge}$
respectively). We will not have to refer to any metric in $\cal F$
so we only need a single Hodge dual symbol $*$. In all relevant
cases --such as (\ref{028})-- both $\wedge$ and ${\wedge
\kern-.57em \wedge}$ appear so, to make notation lighter, we will
only write a ${\wedge \kern-.57em \wedge}$ symbol\footnote{This
could have been avoided by explicitly writing space-time indices;
we feel, however that this would unnecessarily complicate the
notation.}. Notice, also, that  $d$ and $\delta$ are defined in
$\real^4$ but the integral in (\ref{028}) is three-dimensional
and, hence, we cannot ``integrate by parts''. The fact that
$dJ=0$ on solutions allows us to take any space-like slice of
$\real^4=\real^3\times \real$. Notice that this implies that even
though $A(x)$ depends on both spatial and time variables
$\Omega_{{\cal S}}$ --the restriction of $\Omega_{\cal F}$ to
${\cal S}\subset {\cal F}$-- is time independent (we will choose
inertial coordinates $(\vec{x},t)$); in a sense, $\Omega_{{\cal
S}}$ depends on equivalence classes under evolution of initial
data for physical field configurations.

In order to explicitly write $\Omega_{{\cal S}}$ in terms of
solutions we write
\begin{eqnarray*}
A&=&(\gamma^q+d\Lambda^q)e_q+(\gamma^p+\delta \Theta^p)e_p+
(\gamma^r+d\delta\gamma^r_{\scriptscriptstyle H\!D})e_r\\
dA&=&d\gamma^qe_q+(d\gamma^p+d\delta \Theta^p)e_p+d\gamma^re_r\\
\delta A&=&(\delta \gamma^q+\delta d \Lambda^q)e_q+\delta
\gamma^pe_p+ (\delta \gamma^r+\delta d\delta
\gamma^r_{\scriptscriptstyle H\!D})e_r\quad.
\end{eqnarray*}

    A tedious but straightforward computation (described in
Appendix D) gives \vspace*{-0.5cm}
 $$ \Omega_{{\cal
S}}=-2is!\int_{\real^3}\frac{d^3\vec{k}}{w}\dd
\bar{\beta}^\texttt{t}_{i_1\cdots i_s}(\vec{k})\w P\dd
\beta^{i_1\cdots i_s}(\vec{k})\hspace{6.3cm}$$ \vspace*{-0.5cm}$$
+\frac{2iss!}{s-1}\int_{\real^3} \frac{d^3\vec{k}}{w}\dd
\bar{a}^\texttt{t}_{i_1\cdots i_{s-2}}(\vec{k})\w Q \dd
a_{i_1\cdots i_{s-2}}(\vec{k})+\hspace{6.3cm}$$\vspace*{-0.5cm}
\begin{equation} + 2iss!\int_{\real^3}\frac{d^3\vec{k}}{w}\dd
\bar{b}_{i_1\cdots i_{s-1}}^re_r^\texttt{t}(\vec{k})\w P\dd
b^{i_1\cdots
i_{s-1}}(\vec{k})+\label{030}\hspace{5.8cm}\end{equation}\vspace*{-0.5cm}
$$ +is!\int_{\real^3}\frac{d^3\vec{k}}{w}\left[\dd
\bar{\sigma}^r_{i_1\cdots i_{s-1}}(\vec{k}) e^\texttt{t}_r\w P\dd
b^{i_1\cdots i_{s-1}}(\vec{k})- \dd \sigma^r_{i_1\cdots
i_{s-1}}(\vec{k}) e^\texttt{t}_r\w P\dd \bar{b}^{i_1\cdots
i_{s-1}}(\vec{k}) \right]-$$ \vspace*{-0.5cm}$$
-2s!\int_{\real^3}\frac{d^3\vec{k}}{w}\left[
\dd\bar{\alpha}^r_{i_1\cdots i_{s-1}}(\vec{k})e^\texttt{t}_r\w
P\dd b^{i_1\cdots i_{s-1}}(\vec{k}) + \dd \alpha^r_{i_1\cdots
i_{s-1}}(\vec{k})e^\texttt{t}_r\w P\dd \bar{b}^{i_1\cdots
i_{s-1}}(\vec{k}) \right]$$ for\footnote{We use the convention
that an index running from $i_1$ to $i_0$ represents the absence
of an index; if the index runs from $i_1$ to $i_{-k}$ for some
positive integer $k$ then the field itself is zero.} $0\leq s\leq
4$ (in the case $s=1$ the second term in (\ref{030}) is absent
from $\Omega_{{\cal S}}$ and for $s=3$ the first is zero because
$\beta_{i_1i_2i_3}$ is transversal and hence, zero). It is
important to remember that the objects appearing in (\ref{030})
are subject to the algebraic constraints given by (\ref{027}).

    Several comments are in order now.

i) $\Omega_{{\cal S}}$ given  by (\ref{030}) is {\it real};
although we will not write it here in terms of the real and
imaginary parts of the fields.

ii) $\Omega_{{\cal S}}$ is explicitly time independent as expected
from general arguments on the time independence of the symplectic
form. Notice that this comes about through rather non-trivial
cancellations of terms and the explicit use of the field equations
as described in the Appendix D.

iii) The functions $\Lambda^q$ and $\Theta^p$ appearing in the
solutions to the field equations label gauge transformations in
the $p$ and $q$ sectors. We have not explicitly solved the
algebraic constraint. Although one would have to do it in order
to completely identify  the physical degrees of freedom, we do not
need to do it for the purpose of this paper as shown below.

iv) When $s=1$ the second term  in (\ref{030}) is absent, so it
seems that no dependence in $Q$ remains, however this is not the
case because $Q$ enters indirectly through the definition of
$e_r$ and the algebraic equation (\ref{027}).

v) The gauge symmetries of the model can be traced back to the
action (\ref{003}) by writing
\begin{eqnarray*}
dA^\texttt{t}\wedge *PdA&=&d(A^qe^\texttt{t}_q+A^re^\texttt{t}_r)\wedge*Pd(A^qe_q+A^re_r)\\
\delta A^\texttt{t}\wedge*Q\delta
A&=&\delta(A^pe^\texttt{t}_p+A^re^\texttt{t}_r)\wedge*Q\delta(A^pe_p+A^re_r)\quad,
\end{eqnarray*}
so that $A^q$ only appears in $dA^\texttt{t}\wedge *PdA$ and
$A^p$ in $\delta A^\texttt{t}\wedge *Q\delta A$ and the action is
invariant under
\begin{eqnarray*}
A^q\mapsto A^q+d\Lambda^q\quad;\quad A^p\mapsto A^p+\delta
\Theta^p\quad.
\end{eqnarray*}
There is no gauge symmetry in the $r$ sector.

Once we have $\Omega_{\cal S}$ in the physical phase space we can
easily compute the energy-momentum tensor and the angular
momentum. This will allow us to label physical states by their
helicities and find out which conditions $P$ and $Q$ must satisfy
in order to define a consistent theory in the sense described
above.

\section{Energy Momentum, Angular Momentum, and Consistency.}

    After finding the gauge transformations and physical degrees of
freedom described by the action (\ref{003}) we want to choose $P$
and $Q$ leading to a consistent theory. We also want to
find out the helicities of the physical states (they are all,
obviously, massless) to completely characterize them. In order to
do this we need to obtain the energy-momentum and the
angular momentum of the physical modes appearing in (\ref{030})
(after solving for the algebraic constraint). This can be done in
a rather convenient way by using $\Omega_{\cal S}$. This
symplectic structure is invariant under all the symmetries of the
theory because $\Omega_{\cal F}$ is. As is well known the
degenerate directions of the symplectic form give us the gauge
symmetries present in the model. Also if it is invariant under a
group of transformations and we take a vector $V$ tangent to an
orbit of this group it is straightforward to prove \cite{CR1,CR2},
that locally $i_V\Omega_{\cal S}=\dd H$, where $i_V\Omega_{{\cal
S}}$ denotes the contraction of $V$ and $\Omega_{{\cal S}}$, and
the quantity $H$ is the generator of the symmetry transformation
corresponding to $V$. If the action is Poincar\'e invariant we
can obtain in this way the energy-momentum and the angular
momentum (with the right symmetries in their tensor indices) by
computing $i_V\Omega_{{\cal S}}$ for vectors $V$ describing
translations and Lorentz transformations and writing the result
as $\dd H$.

    Space-time translations are given by
\begin{eqnarray*}
x^0 \equiv t\mapsto t+\tau^0\quad;\quad x^i\equiv \vec{x}\mapsto
\vec{x}+\vec{\tau}\quad,
\end{eqnarray*}
they define a vector field in the space of solutions  given by
the formal expression
\begin{eqnarray*}
V_T=\int_{\real^4}d^4 x
\Delta_T\Phi(\vec{x},x^0)\frac{\delta\quad\quad}{\delta\Phi(\vec{x},x^0)}
\end{eqnarray*}
where $\Phi(\vec{x},x^0)$ labels the solutions to the field
equations , $\frac{\delta\quad\,\,}{\delta\Phi(\vec{x},x^0)} $
denotes the functional derivative and $\Delta_T\Phi(\vec{x},x^0) $
is the translation on $\Phi(\vec{x},x^0)$ with parameters
$(\vec{\tau},\tau^0)$. The Fourier transforms of
$\Delta_T\Phi(\vec{x},x^0)$ for the fields appearing in (\ref{030})
are
\begin{eqnarray*}
\Delta_T\beta_{i_1\cdots
i_s}(\vec{k})&=&i\tau^ak_a\beta_{i_1\cdots i_s}(\vec{k})\\
\Delta_T a_{i_1\cdots i_{s-1}}(\vec{k})&=&i\tau^ak_aa_{i_1\cdots i_{s-1}}(\vec{k})\\
\Delta_T b_{i_1\cdots i_{s-1}}(\vec{k})&=&i\tau^ak_ab_{i_1\cdots
i_{s-1}}(\vec{k}) \\
\Delta_T \sigma_{i_1\cdots i_{s-1}}(\vec{k})&=&i\tau^ak_a\sigma_{i_1\cdots i_{s-1}}(\vec{k})\\
\Delta_T\alpha^p_{i_1\cdots
i_{s-1}}(\vec{k})&=&i\tau^ak_a\alpha^p_{i_1\cdots
i_{s-1}}(\vec{k})\\
\Delta_T\alpha^r_{i_1\cdots
i_{s-1}}(\vec{k})&=&i\tau^ak_a\alpha^r_{i_1\cdots
i_{s-1}}(\vec{k}) +\tau^0w\sigma^r_{i_1\cdots
i_{s-1}}(\vec{k})\quad,
\end{eqnarray*}
with $k^0\equiv w$. We do not need the transformations for
$\Lambda^p$, $\Theta^q$. Notice that $\alpha^r_{i_1\cdots i_{s-1}}(\vec{k})$ transforms in
a rather peculiar way due to the terms $wt\sigma^r_{i_1\cdots
i_{s-1}}(\vec{k})$ and $wt\bar{\sigma}^r_{i_1\cdots
i_{s-1}}(-\vec{k})$ in (\ref{025bis}). For the fields appearing in
$\Omega_{{\cal S}}$ we have
\begin{eqnarray*}
i_{V_T}\dd \beta_{i_1\cdots
i_s}(\vec{k})&=&i\tau^ak_a\beta_{i_1\cdots i_s}(\vec{k})\\
i_{V_T}\dd \sigma_{i_1\cdots i_{s-1}}(\vec{k})&=&i\tau^ak_a \sigma_{i_1\cdots i_{s-1}}(\vec{k})\\
i_{V_T}\dd a_{i_1\cdots i_{s-2}}(\vec{k})&=&i\tau^ak_aa_{i_1\cdots
i_{s-2}}(\vec{k})\\
i_{V_T}\dd \alpha^r_{i_1\cdots
i_{s-1}}(\vec{k})&=&i\tau^ak_a\alpha^r_{i_1\cdots
i_{s-1}}(\vec{k})
+\tau^0w\sigma^r_{i_1\cdots i_{s-1}}(\vec{k})\\
i_{V_T}\dd b_{i_1\cdots i_{s-1}}(\vec{k})&=&i\tau^ak_a
b_{i_1\cdots i_{s-1}}(\vec{k})\quad,
\end{eqnarray*}
and hence the energy momentum can be obtained as
\begin{equation}
i_{V_T}\Omega=\tau^a\dd P_a=\hspace{11.1cm} \label{035}
\end{equation}
\vspace*{-8mm} $$\dd\bigg\{
\int_{\real^3}\frac{d^3\vec{k}}{w}\bigg[ -2s!
(\tau^ak_a)\bar{\beta}^\texttt{t}_{i_1\cdots i_s}(\vec{k})P
\beta^{i_1\cdots i_s}(\vec{k}) +\frac{2ss!}{s-1}(\tau^ak_a)
\bar{a}^\texttt{t}_{i_1\cdots i_{s-2}}(\vec{k})Q a^{i_1\cdots
i_{s-2}}(\vec{k})+$$ \vspace*{-10mm}
\begin{eqnarray}
& & +2ss!(\tau^ak_a)\bar{b}^r_{i_1\cdots i_{s-1}}(\vec{k})
e_r^\texttt{t}
P b^{i_1\cdots i_{s-1}}(\vec{k})+\nonumber\\
& & +s!(\tau^ak_a)\bar{\sigma}^r_{i_1\cdots
i_{s-1}}(\vec{k})e^\texttt{t}_rP b^{i_1\cdots i_{s-1}}(\vec{k})
+s!(\tau^ak_a) \sigma^r_{i_1\cdots
i_{s-1}}(\vec{k})e^\texttt{t}_rP\bar{b}^{i_1\cdots i_{s-1}}(\vec{k})+\nonumber\\
& & +2is!(\tau^ak_a)\bar{\alpha}^r_{i_1\cdots
i_{s-1}}(\vec{k})e^\texttt{t}_rP b^{i_1\cdots i_{s-1}}(\vec{k})
+2is!(\tau^ak_a) \alpha^r_{i_1\cdots
i_{s-1}}(\vec{k})e^\texttt{t}_rP\bar{b}^{i_1\cdots i_{s-1}}(\vec{k})-\nonumber\\
& &- s!(\tau^0 w)\bar{\sigma}^r_{i_1\cdots
i_{s-1}}(\vec{k})e^\texttt{t}_rP b^{i_1\cdots
i_{s-1}}(\vec{k})-s!(\tau^0w)\sigma^r_{i_1\cdots
i_{s-1}}(\vec{k})e^\texttt{t}_rP\bar{b}^{i_1\cdots
i_{s-1}}(\vec{k})\bigg]\bigg\}.\nonumber
\end{eqnarray}
In view of (\ref{035}) it must be pointed out that the appearance
of $\tau^0$ does not spoil the Lorentz covariance of $P_a$
because $\alpha^r_{i_1\cdots i_{s-1}}(\vec{k})$ have
transformation laws under space-time translations\footnote{These
objects have also unusual transformation laws under Lorentz
boosts.} that involve $\tau^0$. Also notice that the second term
is absent when $s=1$ and the first one is zero when $s=3$. We
consider now spatial rotations in order to identify the
helicities of the physical states described by the action
(\ref{003}). To this end we need
\begin{eqnarray*}
i_{V_R}\dd \beta_{i_1\cdots i_s}(\vec{k})&=&
-s\varepsilon_{[i_1|jk}\Lambda_j
    \beta_{k|i_2\cdots i_s]}(\vec{k})-
    \varepsilon_{jkl}k_j
    \Lambda_k\frac{\partial\beta_{i_1\cdots i_s}}{\partial k_l}(\vec{k})  \\
i_{V_R}\dd a_{i_1\cdots i_{s-2}}(\vec{k})&=&
-(s-2)\varepsilon_{[i_1|jk}\Lambda_j
    a_{k|i_2\cdots i_{s-2}]}(\vec{k})-
    \varepsilon_{jkl}k_j
    \Lambda_k\frac{\partial a_{i_1\cdots i_{s-2}}}{\partial k_l}(\vec{k})
\\
i_{V_R}\dd b_{i_1\cdots i_{s-1}}(\vec{k})&=&
-(s-1)\varepsilon_{[i_1|jk}\Lambda_j
    b_{k|i_2\cdots i_{s-1}]}(\vec{k})-
    \varepsilon_{jkl}k_j
    \Lambda_k\frac{\partial b_{i_1\cdots i_{s-1}}}{\partial k_l}(\vec{k})
\\
i_{V_R}\dd \sigma_{i_1\cdots i_{s-1}}(\vec{k})&=&
-(s-1)\varepsilon_{[i_1|jk}\Lambda_j
    \sigma_{k|i_2\cdots i_{s-1}]}(\vec{k})-
    \varepsilon_{jkl}k_j
    \Lambda_k\frac{\partial\sigma_{i_1\cdots i_{s-1}}}{\partial k_l}(\vec{k})  \\
i_{V_R}\dd \alpha_{i_1\cdots
i_{s-1}}(\vec{k})&=&-(s-1)\varepsilon_{[i_1|jk}\Lambda_j
    \alpha_{k|i_2\cdots i_{s-1}]}(\vec{k})-
    \varepsilon_{jkl}k_j
    \Lambda_k\frac{\partial\alpha_{i_1\cdots i_{s-1}}}{\partial k_l}(\vec{k})
\end{eqnarray*}
and hence
\begin{equation}
i_{V_R}\Omega_{{\cal S}}=\varepsilon_{ijk}\Lambda_i\dd
J_{jk}=\hspace{10cm}\label{037}
\end{equation}
\vspace{-7mm} $$\dd \bigg\{
\varepsilon_{ijk}\Lambda_i\int_{\real^3}\frac{d^3\vec{k}}{w}\bigg[
-2is!k_j\frac{\partial\bar{\beta}^\texttt{t}_{i_1\cdots
i_s}}{\partial k_k}(\vec{k})P \beta_{i_1\cdots
i_s}(\vec{k})+\hspace{5cm}$$ \vspace{-5mm}
$$+\frac{2iss!}{s-1}k_j\frac{\partial
\bar{a}^\texttt{t}_{i_1\cdots i_{s-2}}}{\partial k_k }(\vec{k})Q
a_{i_1\cdots i_{s-2}}(\vec{k})+2iss!k_j\frac{\partial
\bar{b}^\texttt{t}_{i_1\cdots i_{s-1}}}{\partial
 k_k}(\vec{k})P b_{i_1\cdots i_{s-1}}(\vec{k})-$$
 \vspace{-5mm}
$$-is!k_j\left( \frac{\partial \bar{\sigma}^r_{i_1\cdots
i_{s-1}}}{\partial k_k}(\vec{k})e^\texttt{t}_r P b_{i_1\cdots
i_{s-1}}(\vec{k}) - \frac{\partial \sigma^r_{i_1\cdots
i_{s-1}}}{\partial k_k}(\vec{k})e^\texttt{t}_rP \bar{b}_{i_1\cdots
i_{s-1}}(\vec{k}) \right)+$$ \vspace{-5mm}
$$+2s!k_j\left(\frac{\partial\bar{\alpha}^r_{i_1\cdots
i_{s-1}}}{\partial k_k}(\vec{k})e^\texttt{t}_rP b_{i_1\cdots
i_{s-1}}(\vec{k}) + \frac{\partial \alpha^r_{i_1\cdots
i_{s-1}}}{\partial k_k}(\vec{k})e^\texttt{t}_rP \bar{b}_{i_1\cdots
i_{s-1}}(\vec{k}) \right)+$$ \vspace{-5mm}
$$+2iss!\bar{\beta}^\texttt{t}_{ji_2\cdots i_s}(\vec{k})P
\beta_{ki_2\cdots
 i_s}(\vec{k})-\frac{2i(s-2)s s!}{s-1}\bar{a}^\texttt{t}_{ji_2\cdots
 i_{s-2}}(\vec{k})Q a_{ki_2\cdots i_{s-2}}(\vec{k})-$$
$$-2i(s-1)ss!\bar{b}^r_{j i_2\cdots
i_{s-1}}(\vec{k})e_r^\texttt{t}P b_{ki_2\cdots
 i_{s-2}}(\vec{k})-\hspace{5.6cm}$$
$$\quad-i(s-1)s!\left( \bar{\sigma}^{r}_{ji_2\cdots
i_{s-1}}(\vec{k})e^\texttt{t}_rP b_{ki_2\cdots
i_{s-1}}(\vec{k})-\sigma^{r}_{ji_2\cdots
i_{s-1}}(\vec{k})e^\texttt{t}_rP \bar{b}_{ki_2\cdots
i_{s-1}}(\vec{k})\right)+$$
$$\quad\quad+2(s-1)s!\left(
\bar{\alpha}^{r}_{ji_2\cdots i_{s-1}}(\vec{k})e^\texttt{t}_rP
b_{ki_2\cdots i_{s-1}}(\vec{k})+ \alpha^{r}_{ji_2\cdots
i_{s-1}}(\vec{k})e^\texttt{t}_rP \bar{b}_{ki_2\cdots
i_{s-1}}(\vec{k}) \right)\bigg]\bigg\},$$ we will use this later
in order to identify the helicities of the physical modes for
choices of $P$ and $Q$ leading to consistent models.

    In the following we will find the conditions that $P$ and
$Q$ must satisfy in order to define a  consistent theory. The
main condition that we will impose is the semi-boundedness of the
energy. We need this to ensure that, after coupling the fields to
some others or with themselves via self-interaction terms, we
have a stable theory. If we look at the energy-momentum given by
(\ref{035}) we see that $\beta_{i_1\cdots i_s}(\vec{k})$ and
$a_{i_1\cdots i_{s-2}}(\vec{k})$ are decoupled from the remaining
modes so, to have a positive definite or semi-definite energy
both $P$ and $Q$, when present in (\ref{035}), must be definite or
semi-definite. Notice that in the case $s=1$ (1-form fields) the
term involving $a_{i_1\cdots i_{s-2}}$ is absent and hence we
only have a condition on $P$; if $s=2$ we have conditions both on
$P$ and $Q$ and, finally, if $s=3$ we only have conditions on
$Q$. The remaining terms in (\ref{035}) are proportional to
$e_r$. From the fact that each of them is also proportional to
$Pb$ it is very easy to prove that none of the terms involving
$b$, $\alpha$, and $\sigma$ can be zero for non zero values of
the fields if the $e_r$ sector is present. This is so because the
projection of ${\rm Im}\, P$ on $e_r$  is always non-zero because
${\rm Im}\, P$ is orthogonal to ${\rm Ker}\, P$ ($P$ is a
symmetric matrix) and $e_r$ is orthogonal to ${\rm Ker}\, P$. We
see then that $\alpha$, $b$, and $\sigma$ give a non-zero
contribution to the energy. However, the quadratic form that
defines the energy has some zeroes in its diagonal, in particular
there are no $\bar{\alpha}^r$-$\,\alpha^r$ terms. This fact is
independent of the algebraic constraint because it does not
involve $\alpha^r$. A well known result in linear algebra (see
Appendix E) states that, under the previous conditions, a
quadratic form with a zero in its main diagonal can never be
neither definite nor semi-definite. We conclude
that\footnote{Notice that the effect of not having a $e_r$ sector
can be taken into account by setting $e_r=0$ in the previous
formulas; in such case a basis of $\real^N$ is spanned only by
$e_p$ and $e_q$.} if $e_r\neq 0$ the energy cannot be
semi-bounded and hence the action (\ref{035}) leads to an
inconsistent theory.

We consider now the case $e_r=0$ for which only the terms
containing $\beta$ and $a$ remain. Clearly it suffices now to
choose $P$ and $Q$ (in those cases in which the corresponding
terms are present in (\ref{035})) to be definite or
semi-definite. The question to answer at this point is whether
these models can be non-trivial in the sense that no linear
transformation  of the fields --we want to preserve the quadratic
character of the action-- takes the action (\ref{003}) to the form
\begin{eqnarray}
\int_{\real^4}
\left[
 \sum_{ g= 1}^{n_d} \sigma_g dA_g\wedge * dA_g +
\sum_{g= n_d+1}^{n_d+n_\delta}\sigma_g \delta A_g\wedge * \delta A_g \right]\quad,
\label{038}
\end{eqnarray}
with $\sigma_g=\pm 1$ (and always positive whenever the
corresponding piece describes local degrees of freedom). Notice
that, for example, in the case $s=1$ (\ref{038}) is the sum of several
Maxwell actions and $(\delta A)^2$ terms that carry no
degrees of freedom in a Minkowskian space-time.

    Let us take  $P$ and $Q$ such that\footnote{Remember
that ${\rm Ker}\, P\cap {\rm Ker}\, Q=\{0\}$
so that the sum is indeed a direct sum of vector subspaces of
$\real^N$. Also ${\rm Ker}\, P$ and ${\rm Ker}\, Q$ need not be
mutually orthogonal.} $\real^N={\rm Ker}\, P\oplus {\rm Ker}\,
Q$ so that $e_r=0$. By means of a linear
transformation we can take  $P$ to a diagonal form with ($\dim
{\rm Ker}\, P$)-elements  equal to one and the rest equal to
zero. Let us restrict us for the moment to the cases $s=1$ or $s=2$
and suppose then that $P$ is positive semi-definite. We can
then write in block form as
\begin{eqnarray}
P=\left[\begin{array}{cc}1&0\\0&0\end{array}\right]\quad.\label{039}
\end{eqnarray}
The most general linear, non singular, field redefinition leaving
it invariant is given by the matrix
\begin{eqnarray}
R=\left[\begin{array}{cc}\alpha&\alpha_1\\0&\alpha_2\end{array}\right]
\label{040}
\end{eqnarray}
with $\alpha$ orthogonal and $\alpha_2$ non-singular. If we write
$Q$, with the  same block dimensions of (\ref{039}) as
\begin{eqnarray*}
Q=\left[\begin{array}{cc}a&b\\b^\texttt{t}&c\end{array}\right]\quad,
\end{eqnarray*}
it transforms under (\ref{040}) according to
\begin{eqnarray}
R Q R^{\texttt{t}}
=\left[\begin{array}{cc}
(\alpha a +\alpha_1b^\texttt{t})\alpha^\texttt{t}+(\alpha b+
\alpha_1c^\texttt{t})\alpha_1^\texttt{t}&(\alpha b+\alpha_1 c)\alpha_2^\texttt{t}
\\
\alpha_2(\alpha b+\alpha_1 c)^\texttt{t}&\alpha_2 c\alpha_2^\texttt{t}
\end{array}\right]
\label{042}\end{eqnarray}
Let us suppose now that $c$ is non singular as a ($\dim{\rm
Ker}\, P)\times (\dim{\rm Ker}\, P)$ matrix; then,
$\alpha_1=-\alpha b c^{-1}$ would render the non-diagonal  blocks
of  (\ref{042}) equal to zero. Furthermore, $\alpha$ and $\alpha_2$ can be
chosen in such a way that (\ref{042}) is diagonal; in fact $c$ can be
diagonalized by an othonormal $\alpha_2$. Taking into account
that ${\rm rank} Q=\dim{\rm Ker}\, P={\rm rank} c$ which implies
$(a-bc^{-1}b^\texttt{t})=0$ we find $(\alpha
a+\alpha_1b^\texttt{t})\alpha^\texttt{t}=\alpha(a-bc^{-1}b^\texttt{t})\alpha^\texttt{t}=0$.
We see that by taking a non-singular $c$ we can transform the action (\ref{003})
into (\ref{038}) by a non-singular linear field redefinition.

    Let us suppose now that $c$ is singular; in this case,
one can find non-zero vectors $\rho_c\in\real^{\dim{\rm Ker}\,
P}$ such that $c\rho_c=0$ so if $b\rho_c\neq 0$  there is no way
to mutually diagonalize both  $P$ and $Q$ because the
off-diagonal blocks in (\ref{042}) would be non-zero. If we look at
${\rm Ker}\, Q$ we have
\begin{eqnarray}
\left[
\begin{array}{cc}
    a& b\\b^\texttt{t}&c
\end{array}
\right]
\left[
\begin{array}{c}
    y\\ x
\end{array}
\right]
=
\left[\begin{array}{c}0\\0\end{array}\right]\Rightarrow \left.\begin{array}{l} ay+bx=0\\
b^\texttt{t}y+cx=0\end{array}\right\}\quad.\label{043}
\end{eqnarray}
The second equation in (\ref{043}) implies
$\rho_c^\texttt{t}b^\texttt{t}y=0$ which tells us that the number
of independent $y$ vectors is strictly smaller than $N-\dim{\rm
Ker}\, P$. This means that is impossible to have a basis for
$\real^N$ built only with vectors belonging to ${\rm Ker}\, P$
and ${\rm Ker}\, Q$; i.e. we would have some vector $e_r$
different from zero. We conclude  then that the requirement that
$e_r=0$ forces us to  choose $P$ and $Q$ in such a way that they
can be simultaneously diagonalized leading to a ``trivial''
action of the type (\ref{038}).

Finally, notice that for a singular $c$ such that for every
$\rho_c\in{\rm Ker}\, c$ we have $b\rho_c=0$ we can still make the
non-diagonal blocks equal to zero and from the fact that a
symmetric singular matrix admits a symmetric
pseudoinverse\footnote{$Cx=y$ implies $PCP^\texttt{t}Px=Py$ and
hence $DPx=Py$ where $D$ is diagonal. We can write $Px=D^{-}Py$
with $D^{-}$ consisting on the inverses of the non-zero
eigenvalues of $C$ so that $x=P^\texttt{t}D^{-}P y$ which proves
that there always exist a symmetric pseudoinverse.} we can easily
prove that $Q$ and $P$ are simultaneously diagonalizable.

The case $s=3$ can be analyzed
by  following  the same lines just by switching the roles of $P$
and $Q$.

In view of the previous discussion we see that, whenever the
energy is positive definite or semi-definite --something that
happens only when $e_r=0$-- the action can be transformed by means
of a non-singular linear field redefinition into an action of the
type (\ref{038}).  When $e_r=0$ it is  very easy to analyze the
helicities of the physical modes as only the terms involving
$\beta$ and $a$ should be taken into account. The coefficients of
these terms are the same as those of the corresponding terms in
the symplectic form. (\ref{030}). For $s=1$ we have helicities
$\pm 1$, for $s=2$  we have scalars (notice that for $s=2$ and
$e_r=0$ the spin part of (\ref{037}) is zero) and for $s=3$ we
find again helicities $\pm 1$.

Another important consequence of the previous arguments is the
impossibility of adding a kinetic term written in terms of
1-forms to the tetrad gravity action (\ref{001}). This is so
because $e_I$ and $A_I^{\;\;J}$ transform as an internal vector
and a connection respectively. The connection can be identified
as a field coming from  the first term of (\ref{038}) after  a
suitable deformation of the Yang-Mills type. However, the
transformation law  of $e_I$ is such that it cannot be derived
neither from a connection in the first term ($P$-term) of
(\ref{038}) nor a field in the $Q$-term.

It is very important to understand the key role played by the
condition that the energy be definite or semi-definite in this
respect. If one  relaxes this condition, it is actually possible
to find actions with well defined propagators after the gauge
fixing and with diff-invariant interaction terms very similar  to
the tetrad action  (\ref{001}). Let us consider for example
\begin{eqnarray}
S=\int_{\real^4} \left[\nabla e^I\wedge *\nabla e_I +F^I\wedge
*\nabla e_I+De^I\wedge * D e_I+\varepsilon_{IJK}e^I\wedge e^J\wedge
F^K\right]\quad,
\label{044}
\end{eqnarray}
where now $I=1,2,3$ label (internal) $SO(3)$ indices,
$\varepsilon_{IJK}$ in the $3$-dimensional totally antisymmetric
object, $\nabla e^I= d e^I+[A,e]^I$, $F^I=2dA^I+[A,A]^I$,
$De^I=\delta e^I+[i_A,e]^I$. This action has a quadratic term
leading to a well defined propagator after gauge fixing, is
power-counting renormalizable and has, as an interacting term, the
action for the Husain-Kucha\v r model\footnote{The Husain-Kucha\v
r model \cite{HK} is a toy model for general relativity that has 3 local
degrees of freedom per space point due to the absent of a scalar
constraint in its Hamiltonian formulation.} which mimics a term
of the type defined by (\ref{001}). It also has the property that there
are no regular field redefinitions that allow us to remove the
$F^I\wedge \nabla e_I$ term to make the kinetic term diagonal.
It is only the fact that the energy is not semi-bounded
that leads to inconsistencies. This is reminiscent, but not equal,
to the well known behavior of higher derivative theories of
gravity. The reader may argue that this is to be
expected due to the presence of the $D e_I\wedge *De^I$ term in (\ref{044})
as terms similar to  this are known to spoil, for example, the
familiar Maxwell action. Though, at the end of the day, this
happens to be the case, the reasons, as shown above, are not
obvious. In fact, there are actions involving $\delta e$
that are consistent (albeit trivial).

\section{ Conclusions and Comments.}

The first conclusion of the paper is that the free actions
considered in the literature \cite{YM}-\cite{HennK}  to
study consistent interactions between gauge fields are not the
most general ones in a precise sense. In these papers the
starting point is always a Maxwell-type of action for $s$-form
fields. In some instances; for example 1-forms, one can somehow
extend some of the results already known for 1-forms to
3-forms fields because  the second term in (\ref{002}) can be considered
as a Maxwell action for 3-forms. If one does not allow
interactions  between the ``$P$-sector'' and the ``$Q$-sector''
the results in the literature already apply to this case.
Notice, however, that the 2-form case is different in this
respect  as both terms in (\ref{003}) can be interpreted in terms
of 2-forms as disjoint sectors  of  a Maxwell-like action.

    Our result is useful  because it  gives  general
free actions that should be taken as starting points to the
study of their consistent interactions so we hope that a deeper
knowledge about the uniqueness of Yang-Mills can be achieved by
considering their deformations.

\bigskip

    We want to remark that the use of covariant symplectic techniques
for quadratic theories is very convenient in several respects.
First of all it is much simpler than the use of the familiar
Dirac formalism. There, the appearance of successive layers of
secondary constraints depending  on the algebraic properties of
the $P$ and $Q$ matrices requires tedious computations to
disentangle  the structure of the phase space, constraints, and
gauge symmetries. Here, as also shown in \cite{Nos}, the
possibility of explicitly solving the field equations (via
Fourier transform and after a suitable $3+1$ splitting) allows us
to use the covariant symplectic formalism to describe the
physical degrees of freedom and symmetries of the model.  As we
have seen, it is actually easy to find the full symmetries of the
Lagrangian (rather than the 3-dimensional version provided by the
constraints in the Hamiltonian formalism). We have seen also that
these symplectic techniques help in the derivation of the
energy-momentum and angular momentum, key ingredients to study
the particle content and consistency of the actions considered in
this paper.

    The use of these covariant symplectic techniques offers
the possibility of studying in a very systematic way whole families
of quadratic actions for different kinds of fields. Our point of
view is that the only way to build a perturbatively consistent
theory is to completely understand the quadratic part
of the action and the subsequent deformations of it. We think that
no systematic study of quadratic  gauge actions has been carried
out to date. This paper and the previous one dealing with the
diff-invariant case, are first steps in the program of
characterizing large sets of gauge quadratic actions. We hope
that interesting theories   may appear in this search hopefully
leading to a new understanding of Yang-Mills and other theories such as
general relativity. For example, as shown in \cite{FEBF}, the Husain-Kucha\v{r} 
model can be described by
coupling two BF Lagrangians with a quadratic part involving cross-terms
with 1 and 2-forms. An open and interesting question is if one can find
a quadratic action with consistent deformations that include the
Husain-Kucha\v{r} model in its BF description. Work in this direction is
in progress.

\bigskip

A second question that has been answered in passing concerns the
impossibility of adding quadratic 1-form terms to the tetrad action for
gravity, even in the presence of a metric background such as Minkowski.
This result goes beyond the negative conclusion of \cite{Nos} where we
showed that no diff-invariant kinetic terms could be consistently
added to the gravitational action. Here we have seen that  the
only consistent actions in terms of 1-forms are just Maxwell
actions and $(\delta A)^2$ actions (that describe no
degrees of freedom in a Minkowski background), so even in the presence
of a background it is impossible to find suitable kinetic
terms.

As emphasized above the requirement that the energy be definite or
semi-definite is crucial. It is also important to realize that
one should not be tempted to believe that the presence of $\delta
A^\texttt{t}\wedge *Q\delta A$ terms trivially leads to an
inconsistent theory; in fact if $P=0$ the action is consistent
albeit trivial. The presence of the $\{e_r\}$ sector and its
detailed structure is the key element to explain why the theory is
inconsistent in many cases.

\section*{Appendix A.}

    As shown in the paper it is, at times, quite useful to
translate from tensor notation to index-free form notation so we
provide in this appendix a dictionary to go from one
representation to the other. This will  also allow us to fix
several conventions needed when writing forms as totally
covariant antisymmetric tensors.

We will write a $s$-form $\omega$ defined on a differentiable
manifold $\cal M$ of dimension $N$ (endowed with coordinates $x^a$) as
\begin{eqnarray*}
\omega(x)=\omega_{a_1\cdots a_s}(x)dx^{a_1}\wedge\cdots\wedge dx^{a_s}
\end{eqnarray*}
with
\begin{eqnarray*}
\omega_{a_1\cdots a_s}=\omega_{[a_1\cdots a_s]}\equiv \frac{1}{s!}
\sum_{\pi\in{\cal S}_s}(-1)^\pi\omega_{\pi(a_1)\cdots \pi(a_s)}\, .\,
{\rm (}\pi\in{\cal S}_s\,{\rm is\, a\, permutation \,of\, order\,} s{\rm ).}
\nonumber
\end{eqnarray*}

The exterior (wedge) product of a $s$-form $\omega$ and a $r$-form
$\xi$ is defined as
\begin{eqnarray*}
\omega\wedge \xi =\omega_{[a_1\cdots a_s}\xi_{b_1\cdots
b_r]}dx^{a_1}\wedge\cdots\wedge dx^{a_s}\wedge
dx^{b_1}\wedge\cdots \wedge dx^{b_r}\quad,
\end{eqnarray*}
and satisfies
\begin{eqnarray*}
\omega\wedge \xi&=&(-1)^{sr}\xi\wedge \omega\quad,\\
(\xi\wedge\eta)\wedge \omega&=& \xi\wedge (\eta\wedge \omega)\quad.
\end{eqnarray*}
We define the exterior differential that takes a $s$-form $\omega$ to a
$(s+1)$-form as
\begin{eqnarray*}
d\omega=\partial_{[a_1}\omega_{a_2\cdots a_{s+1}]}dx^{a_1}\wedge\cdots
\wedge dx^{a_{s+1}}\quad,
\end{eqnarray*}
and satisfies
\begin{eqnarray*}
d^2&=&0\\
d(\omega\wedge\xi)&=& d \omega\wedge\xi+(-1)^s\omega\wedge d\xi\quad.
\end{eqnarray*}
In the presence of a non-degenerate metric in $\cal M$ we can
define the Hodge dual of a $s$-form $\omega$ as the $(N-s)$-form
given by
\begin{eqnarray*}
*\omega=\frac{1}{(N-s)!}\frac{1}{\sqrt{|\det g|}}\omega_{b_1\cdots
b_s}\tilde{\eta}^{b_1\cdots b_sc_1\cdots c_{N-s}}g_{a_1c_1}\cdots
g_{a_{N-s}c_{N-s}}dx^{a_1}\wedge\cdots\wedge dx^{a_{N-s}}\, ,
\end{eqnarray*}
where $\tilde{\eta}^{b_1\cdots b_N}$ is the Levi-Civita tensor
density on $\cal M$ defined  to  be, in any coordinate chart, $+1$
for even permutations of the indices and $-1$ for odd
permutations. If $g_{ab}$ has Riemannian signature we have
\begin{eqnarray*}
**\omega=(-1)^{s(N-s)}\omega
\end{eqnarray*}
whereas for Lorentzian signatures we find
\begin{eqnarray*}
**\omega=(-1)^{s(N-s)+1} \omega\quad.
\end{eqnarray*}
We can also define the adjoint exterior differential $\delta$ as
\begin{eqnarray*}
\delta &=&(-1)^{N(s+1)+1}*d*\quad {\rm Riemannian \,\, signature}\\
\delta &=& (-1)^{N(s+1)}*d*\quad \quad{\rm Lorentzian \,\,
signature}.
\end{eqnarray*}
It takes $s$-forms to $(s-1)$-forms according to
\begin{eqnarray*}
\delta\omega=-s\nabla^a\omega_{aa_1\cdots a_{s-1}}d
x^{a_1}\wedge \cdots \wedge d^{a_{s-1}}\quad,
\end{eqnarray*}
where $\nabla$ is the metric compatible, torsion-free, covariant derivative and
 satisfies
\begin{eqnarray*}
\delta^2=0\quad.
\end{eqnarray*}
Finally we define the wave operator
\begin{eqnarray*}
\Box= d\delta +\delta d\quad,
\end{eqnarray*}
that takes $s$-forms to $s$-forms and is given by
\begin{eqnarray*}
\Box \omega=-\nabla_a\nabla^a\omega_{a_1\cdots
a_s}dx^{a_1}\wedge\cdots \wedge x^{a_s};
\end{eqnarray*}
it commutes with both $d$ and $\delta$.

    As already stated in the main text we will refer to the
components of differential forms in a Minkowskian background as follows
\begin{eqnarray*}
\omega=\left\{\begin{array}{c}\omega_{i_1\cdots
i_s}\\
\omega_{0i_1\cdots i_{s-1}}
\end{array}\right\}\quad.
\end{eqnarray*}
With the definition for the Fourier transform given by (\ref{019}) we
have the following useful formulas for the various differential
operators acting on $s$-forms. Let us write a general $s$-form
$\omega$ as
\begin{eqnarray*}
\omega(\vec{k},t)=\left\{\begin{array}{c}
ik_{[i_1}\alpha_{i_2\cdots i_s]}(\vec{k},t) +\beta_{i_1\cdots
i_s}(\vec{k},t)
\\ \\
ik_{[i_1}a_{i_2\cdots i_{s-2}]}(\vec{k},t) +b_{i_1\cdots
i_{s-1}}(\vec{k},t)
\end{array}\right\}
\end{eqnarray*}
with $\alpha_{i_1\cdots i_{s-1}}$, $\beta_{i_1\cdots i_s}$,
$a_{i_1\cdots i_{s-2}}$, and $b_{i_1\cdots i_{s-1}}$ transversal.
We have now (dots represent time derivatives and
$w=+\sqrt{\vec{k}\cdot \vec{k}}\quad$)
\begin{eqnarray*}
d\omega&=&\left\{\begin{array}{c} ik_{[i_1}\beta_{i_2\cdots
i_{s+1}]}\\ \\ik_{[i_1}\left(\frac{1}{s+1}\dot{\alpha}_{i_2\cdots
i_s]}
                -\frac{s}{s+1}b_{i_2\cdots i_s]}\right)
       +\frac{1}{s+1}\dot{\beta}_{i_1\cdots i_s}
   \end{array}\right\}\\
\delta\omega&=&
       \left\{\begin{array}{c} ik_{[i_1}s\dot{a}_{i_2\cdots i_{s-1}]}
                        +
                        w^2\alpha_{i_1\cdots i_{s-1}}
                        +s\dot{b}_{i_1\cdots i_{s-1}}\\ \\
        -\frac{s}{s-1}w^2a_{i_1\cdots i_{s-2}}
        \end{array}\right\}\\
d\delta\omega&=&
       \left\{\begin{array}{c}  ik_{[i_1}\left(s\dot{b}_{i_2\cdots i_s]}
       +w^2\alpha_{i_2\cdots i_s]}\right)\\ \\
                   ik_{[i_1}
        \left(
        \ddot{a}_{i_2\cdots i_{s-1}]}+w^2a_{i_2\cdots i_{s-1}]}
        \right)
         + \ddot{b}_{i_1\cdots i_{s-1}}+\frac{w^2}{s}\dot{\alpha}_{i_1\cdots i_{s-1}}
           \end{array}\right\}\\
  \delta d\omega&=&     \left\{\begin{array}{c}    ik_{[i_1}
        \left(
        \ddot{\alpha}_{i_2\cdots i_s]}-s\dot{b}_{i_2\cdots i_s]}
        \right)
        +\ddot{\beta}_{i_1\cdots i_s}+w^2\beta_{i_1\cdots i_s}\\ \\
        w^2b_{i_1\cdots i_{s-1}}-\frac{w^2}{s}\dot{\alpha}_{i_1\cdots i_s}
        \end{array}\right\}\\
\Box \omega    &=&   \left\{\begin{array}{c} ik_{[i_1}
        \left(
        \ddot{\alpha}_{i_2\cdots i_s]}+w^2\alpha_{i_2\cdots i_s]}
        \right)
  +\ddot{\beta}_{i_1\cdots i_s}+w^2\beta_{i_1\cdots i_s}\\ \\
    ik_{[i_1}
        \left(
        \ddot{a}_{i_2\cdots i_{s-1}]}+w^2a_{i_2\cdots i_{s-1}]}
        \right)
         + \ddot{b}_{i_1\cdots i_{s-1}} +w^2b_{i_1\cdots i_{s-1}}
\end{array}\right\}\quad.
\end{eqnarray*}
We will refer to a form satisfying the wave equation $\Box
\gamma=0$ as ``harmonic" even though this term usually refers to
forms satisfying a Laplace equation. If $\gamma$ satisfies
$\Box \gamma=0$ the objects $\alpha_{i_1\cdots i_{s-1}}$,
$\beta_{i_1\cdots i_s}$, $a_{i_1\cdots i_{s-2}}$, and
$b_{i_1\cdots i_{s-1}}$ can be parametrized as
\begin{eqnarray}
w\alpha_{i_1\cdots i_{s-1}}(\vec{k},t) &=&
    \alpha_{i_1\cdots i_{s-1}}(\vec{k})e^{-iwt}
    +
   \bar{\alpha}_{i_1\cdots i_{s-1}}(-\vec{k})e^{iwt}\nonumber\\
\beta_{i_1\cdots i_s}(\vec{k},t) &=&
    \beta_{i_1\cdots i_s}(\vec{k})e^{-iwt}
    +
   \bar{\beta}_{i_1\cdots i_s}(-\vec{k})e^{iwt}\label{alfa}\\
wa_{i_1\cdots i_{s-2}}(\vec{k},t) &=&
    a_{i_1\cdots i_{s-2}}(\vec{k})e^{-iwt}
    +
   \bar{a}_{i_1\cdots i_{s-2}}(-\vec{k})e^{iwt}\nonumber\\
b_{i_1\cdots i_{s-1}}(\vec{k},t) &=&
    b_{i_1\cdots i_{s-1}}(\vec{k})e^{-iwt}
    +
   \bar{b}_{i_1\cdots i_{s-1}}(-\vec{k})e^{iwt}\nonumber
\end{eqnarray}
For harmonic $s$-forms we have
\begin{eqnarray*}
 d\gamma&=& \left\{\begin{array}{c} ik_{[i_1}
 \left[\beta_{i_2\cdots i_{s+1}]}(\vec{k})e^{-iwt}
 +\bar{\beta}_{i_2\cdots i_{s+1}]}(-\vec{k})e^{iwt}
\right]\\ \\
i k_{[i_1}\left[\frac{1}{s+1} \left(-i\alpha_{i_2\cdots
i_s]}(\vec{k})-sb_{i_2\cdots i_s]}(\vec{k})\right)e^{-iwt}
+\right.\hspace{3cm}\\
+\left.\frac{1}{s+1} \left(i\bar{\alpha}_{i_2\cdots
i_s]}(-\vec{k})-s\bar{b}_{i_2\cdots
i_s]}(-\vec{k})\right)e^{iwt}\right]+\\
\hspace{3cm}-\frac{iw}{s+1}\beta_{i_1\cdots
i_s}(\vec{k})e^{-iwt}+\frac{iw}{s+1}\bar{\beta}_{i_1\cdots
i_s}(-\vec{k})e^{iwt}
              \end{array}\right\}
\end{eqnarray*}
\begin{eqnarray*}
 \delta \gamma &=&      \left\{\begin{array}{c}
 ik_{[i_1}\left[-is a_{i_2\cdots i_{s-1}]}(\vec{k})e^{-iwt}
+is\bar{a}_{i_2\cdots i_{s-1}]}(-\vec{k})e^{iwt}\right]+\hspace{1.5cm}\\
    +\left[w\alpha_{i_1\cdots i_{s-1}}(\vec{k})-iws b_{i_1\cdots
    i_{s-1}}(\vec{k})\right]e^{-iwt}+\\
     \hspace{3cm}+\left[w\bar{\alpha}_{i_1\cdots i_{s-1}}(-\vec{k})
     +iws\bar{b}_{i_1\cdots
    i_{s-1}}(-\vec{k})\right]e^{iwt}\\
    \\  -\frac{s}{s-1}w a_{i_1\cdots i_{s-2}}(\vec{k})e^{-iwt}
    -\frac{s}{s-1}w \bar{a}_{i_1\cdots i_{s-2}}(-\vec{k})e^{iwt}
        \end{array}\right\}
\end{eqnarray*}
\begin{eqnarray*}
d\delta \gamma &=&      \left\{\begin{array}{l}     ik_{[i_1}\left[
             \left(w\alpha_{i_2\cdots i_s]}(\vec{k})-iwsb_{i_2\cdots
        i_s]}(\vec{k})\right)e^{-iwt}+\right.\hspace{3cm}\\
       \hspace{3cm}+\left.\left(w\bar{\alpha}_{i_2\cdots i_s]}(-\vec{k})
       +iws\bar{b}_{i_2\cdots
        i_s]}(-\vec{k})\right)e^{iwt}\right]\\ \\
    -w^2\left[b_{i_1\cdots
    i_{s-1}}(\vec{k})+\frac{i}{s}\alpha_{i_1\cdots
    i_{s-1}}(\vec{k})\right]e^{-iwt}-\\
    \hspace{3cm}- w^2\left[\bar{b}_{i_1\cdots
    i_{s-1}}(-\vec{k})-\frac{i}{s}\bar{\alpha}_{i_1\cdots
    i_{s-1}}(-\vec{k})\right]e^{iwt}
           \end{array}\right\}
\end{eqnarray*}
\begin{eqnarray*}
  \delta d\gamma &=&     \left\{\begin{array}{l}  ik_{[i_1}\left[-
             \left(w\alpha_{i_2\cdots i_s]}(\vec{k})-iwsb_{i_2\cdots
        i_s]}(\vec{k})\right)e^{-iwt}-\right.\\
       \hspace{3cm}\left.-\left(w\bar{\alpha}_{i_2\cdots i_s]}
       (-\vec{k})+iws\bar{b}_{i_2\cdots
        i_s]}(-\vec{k})\right)e^{iwt}\right]\\ \\
    w^2\left[b_{i_1\cdots
    i_{s-1}}(\vec{k})+\frac{i}{s}\alpha_{i_1\cdots
    i_{s-1}}(\vec{k})\right]e^{-iwt}+\\
    \hspace{3cm}+ w^2\left[\bar{b}_{i_1\cdots
    i_{s-1}}(-\vec{k})-\frac{i}{s}\bar{\alpha}_{i_1\cdots
    i_{s-1}}(-\vec{k})\right]e^{iwt}   \end{array}\right\}
\end{eqnarray*}
and trivially
\begin{eqnarray*}
  \Box \gamma &=&     \left\{\begin{array}{c}  0\\0  \end{array}\right\}.
\end{eqnarray*}

\section*{Appendix B. Solutions to $d\Box \omega=0$ and $\delta \Box \omega=0$.}

    The general solution to the system of equations
\begin{eqnarray*}
d\Box \omega&=&0\\
\delta \Box \omega&=&0
\end{eqnarray*}
can be written as $\omega=\gamma+d\delta \gamma_{\scriptscriptstyle {H\!D}}$ where
$\gamma$ is a general solution to $\Box \omega=0$ and
$\gamma_{\scriptscriptstyle {H\!D}}$ is a general
solution to $\Box^2\omega=0$. We prove
in the following that such general solution can be parametrized as
\begin{eqnarray}
\gamma&=&\left\{\begin{array}{c} \beta_{i_1\cdots
i_s}(\vec{k})e^{-iwt}
            +\bar{\beta}_{i_1\cdots i_s}(-\vec{k})e^{iwt}\\ \\
            \frac{i}{w}k_{[i_1}\left[ a_{i_2\cdots i_{s-1}]}(\vec{k})e^{-iwt}+
            \bar{a}_{i_2\cdots i_{s-1}]}(-\vec{k})e^{iwt}\right]+\hspace{2cm}\\
\hspace{3cm}+b_{i_1\cdots i_{s-1}}(\vec{k}) e^{-iwt}
            +\bar{b}_{i_1\cdots i_{s-1}}(-\vec{k})e^{iwt}
\end{array}\right\}\, ,
\label{B2}
\end{eqnarray}
\begin{eqnarray}
d\delta\gamma_{\scriptscriptstyle
{H\!D}}&=&\left\{\begin{array}{l}
   \frac{i}{w}k_{[i_1}\left\{
    \left[\alpha_{i_2\cdots i_s]}(\vec{k})+wt\sigma_{i_2\cdots i_s]}
    (\vec{k})\right]e^{-iwt}+\right.\\
    \hspace{4cm}+\left.
 \left[\bar{\alpha}_{i_2\cdots i_s]}(-\vec{k})+wt\bar{\sigma}_{i_2\cdots i_s]}
 (-\vec{k})\right]e^{iwt}
 \right\}\\  \\
\frac{1}{s}\left[\sigma_{i_1\cdots i_{s-1}}(\vec{k})
    -
    i\alpha_{i_1\cdots i_{s-1}}(\vec{k})-iwt\sigma_{i_1\cdots
    i_{s-1}}(\vec{k})\right]e^{-iwt}+\\
  \quad\quad +
    \frac{1}{s}\left[\bar{\sigma}_{i_1\cdots i_{s-1}}(-\vec{k})
    +
    i\bar{\alpha}_{i_1\cdots i_{s-1}}(-\vec{k})+iwt\bar{\sigma}_{i_1\cdots
    i_{s-1}}(-\vec{k})\right]e^{iwt}
\end{array}\right\}
\label{B2bis}
\end{eqnarray}
To this end it suffices to write a general solution to
$\Box^2\omega=0$, compute $d\delta \omega$, add to it a general
solution to the wave equation, and absorb in a single term those
that may be written as solutions to either equation. A general
harmonic form is given by (\ref{alfa}) and a general solution to
$\Box^2\omega=0$ can be parametrized as
\begin{eqnarray*}
\gamma_{{\scriptscriptstyle H\!D}}=\left\{\begin{array}{l}
    ik_{[i_1}\left\{
\left[\rho_{i_2\cdots
i_s]}(\vec{k})+\frac{t}{w^2}\sigma_{i_2\cdots
i_s]}(\vec{k})\right] e^{iwt}\right.+\\
    \hspace{4.5cm}+\left.
\left[\bar{\rho}_{i_2\cdots
i_s]}(-\vec{k})+\frac{t}{w^2}\bar{\sigma}_{i_2\cdots
i_s]}(-\vec{k})\right]e^{-iwt}\right\}\\
+\left[\mu_{i_1\cdots i_s}(\vec{k})+wt\nu_{i_1\cdots
i_s}(\vec{k})\right]e^{iwt}
    +
\left[\bar{\mu}_{i_1\cdots i_s}(-\vec{k})+wt\bar{\nu}_{i_1\cdots
i_s}(-\vec{k})\right]e^{-iwt}\\
\\
    ik_{[i_1}\left\{
    \left[r_{i_2\cdots i_{s-1}]}(\vec{k})+wts_{i_2\cdots
    i_{s-1}]}(\vec{k})\right]e^{iwt}+\right.
    \\\hspace{4.5cm}+\left.
\left[\bar{r}_{i_2\cdots i_{s-1}]}(-\vec{k})+wt\bar{s}_{i_2\cdots
i_{s-1}]}(-\vec{k})\right]e^{-iwt}\right\}\\
+\left[m_{i_1\cdots i_{s-1}}(\vec{k})+wtn_{i_1\cdots
i_{s-1}}(\vec{k})\right]e^{iwt}+
   \\ \hspace{4.5cm}+
\left[\bar{m}_{i_1\cdots i_{s-1}}(-\vec{k})+wt\bar{n}_{i_1\cdots
i_{s-1}}(-\vec{k})\right]e^{-iwt}\end{array}\right\}
\end{eqnarray*}
and $d\delta$ acting on a general solution to $\Box^2 \omega=0$
is given by
\begin{eqnarray*}
\left\{\begin{array}{l}
    ik_{[i_1}\left\{\left[
     swn_{i_2\cdots i_s]}(\vec{k})-iswm_{i_2\cdots i_s]}(\vec{k})+w^2\rho_{i_2\cdots
    i_s]}(\vec{k})
     -\hspace{2cm}\right.\right.\\
    \hspace{5cm}\left.-isw^2tn_{i_2\cdots i_s]}(\vec{k})+t\sigma_{i_2\cdots
    i_s]}(\vec{k})
    \right]e^{-iwt}+\\
    +\left[
     sw\bar{n}_{i_2\cdots i_s]}(-\vec{k})+isw\bar{m}_{i_2\cdots i_s]}(-\vec{k})
    +w^2\bar{\rho}_{i_2\cdots i_s]}(-\vec{k})
     +\right.\hspace{2cm}\\
    \hspace{5cm}\left.\left.+isw^2t\bar{n}_{i_2\cdots i_s]}(-\vec{k})
    +t\bar{\sigma}_{i_2\cdots i_s]}(-\vec{k})
    \right]e^{iwt}
    \right\}\\ \\ \\
     ik_{[i_1}\left[- 2iw^2s_{i_2\cdots
    i_{s-1}]}(\vec{k})e^{-iwt}+2iw^2\bar{s}_{i_2\cdots
    i_{s-1}]}(-\vec{k})e^{iwt}\right]-\hspace{2cm}\\
    \hspace{6cm}-\left[2iw^2n_{i_1\cdots i_{s-1}}(\vec{k})+w^2m_{i_1\cdots
    i_{s-1}}(\vec{k})-\right.\\
    -\frac{1}{s}\sigma_{i_1\cdots
    i_{s-1}}(\vec{k})+\frac{iw^3}{s}\rho_{i_1\cdots
    i_{s-1}}(\vec{k})+\\
    \hspace{4.7cm}\left.
    +w^3tn_{i_1\cdots i_{s-1}}(\vec{k})+\frac{iwt}{s}\sigma_{i_1\cdots i_{s-1}}(\vec{k})
      \right]e^{-iwt}+\\
      +\left[2iw^2\bar{n}_{i_1\cdots i_{s-1}}(-\vec{k})-w^2\bar{m}_{i_1\cdots
    i_{s-1}}(-\vec{k})+\right.\\
    \hspace{6cm}+\frac{1}{s}\bar{\sigma}_{i_1\cdots
    i_{s-1}}(-\vec{k})+\frac{iw^3}{s}\bar{\rho}_{i_1\cdots
    i_{s-1}}(-\vec{k})-\\
    \left.
    -w^3t\bar{n}_{i_1\cdots i_{s-1}}(-\vec{k})
    +\frac{iwt}{s}\bar{\sigma}_{i_1\cdots i_{s-1}}(-\vec{k})
      \right]e^{iwt}
\end{array}\right\}.
\end{eqnarray*}
Looking at the previous two expressions we see that the terms
involving time dependent exponentials and objects such as
$wte^{iwt}$ appear in the right combinations to allow the field
redefinitions leading us to write (\ref{B2}) and (\ref{B2bis}).

\section*{Appendix C. The algebraic constraints.}

The field equations  are
\begin{eqnarray}
P\delta d A +Qd\delta A=0\quad.
\label{c1}
\end{eqnarray}
The general solution to it  has been obtained by solving a set of
necessary conditions and substituting them back into (\ref{c1}). In this
way we get
\begin{eqnarray*}
P(\delta d\gamma^qe_q+\delta d\gamma^r e_r)+Q(d\delta
\gamma^pe_p+d\delta\gamma^re_r+d\delta d\delta
\gamma^r_{\scriptscriptstyle H\!D}e_r)=0\quad.
\end{eqnarray*}
Taking into account that
\begin{eqnarray*}
\delta d\gamma^q&=& \left\{\begin{array}{c}ik_{[i_1}
    \left(iwsb^q_{i_2\cdots i_s]}(\vec{k})e^{-iwt}
        -iws\bar{b}^q_{i_2\cdots i_s]}(-\vec{k})e^{iwt}\right)\\ \\
        w^2b^q_{i_1\cdots i_{s-1}}(\vec{k})e^{-iwt}
    +
    w^2\bar{b}^q_{i_1\cdots i_{s-1}}(-\vec{k})e^{iwt}
\end{array}\right\}\\
\delta d\gamma^r&=&\left\{\begin{array}{c} ik_{[i_1}
    \left(iwsb^r_{i_2\cdots i_s]}(\vec{k})e^{-iwt}
        -iws\bar{b}^r_{i_2\cdots i_s]}(-\vec{k})e^{iwt}\right)\\ \\
        w^2b^r_{i_1\cdots i_{s-1}}(\vec{k})e^{-iwt}
    +
    w^2\bar{b}^r_{i_1\cdots i_{s-1}}(-\vec{k})e^{iwt}
\end{array}\right\}\\
d\delta \gamma^p&=&\left\{\begin{array}{l} ik_{[i_1}\left(
            w\alpha^p_{i_2\cdots i_s]}(\vec{k})e^{-iwt}
            +
            w\bar{\alpha}^p_{i_2\cdots
            i_s]}(-\vec{k})e^{iwt}\right)\\ \\
            -\frac{iw^2}{s}\alpha^p_{i_1\cdots i_{s-1}}(\vec{k})e^{-iwt}
    +
            \frac{iw^2}{s}\bar{\alpha}^p_{i_1\cdots i_{s-1}}(-\vec{k})e^{iwt}
\end{array}\right\}\\
d\delta \gamma^r&=& \left\{\begin{array}{l} ik_{[i_1}\left(
    -iswb^r_{i_2\cdots i_s]}(\vec{k})e^{-iwt}
+ isw\bar{b}^r_{i_2\cdots i_s]}(-\vec{k})e^{iwt}\right)\\ \\
-w^2b^r_{i_1\cdots i_{s-1}}(\vec{k})e^{-iwt}
    -w^2\bar{b}^r_{i_1\cdots i_{s-1}}(-\vec{k})e^{iwt}
\end{array}\right\}\\
d\delta d\delta\gamma^r_{{\scriptscriptstyle H\!D}}&=&\left\{\begin{array}{l}
    ik_{[i_1}\left(-2iw\sigma^r_{i_2\cdots
    i_s]}(\vec{k})e^{-iwt}+2iw\bar{\sigma}^r_{i_2\cdots
    i_s]}(-\vec{k})e^{iwt}\right)\\ \\
    -\frac{2w^2}{s}\sigma^r_{i_1\cdots
    i_{s-1}}(\vec{k})e^{-iwt}-\frac{2w^2}{s}\bar{\sigma}^r_{i_1\cdots
    i_{s-1}}(-\vec{k})e^{iwt}
\end{array}\right\}\quad,
\end{eqnarray*}
we find the Fourier transform of the spatial part of the constraint
\begin{eqnarray*}
P\left[iswb^q_{i_1\cdots i_{s-1}}(\vec{k}) e_q+iswb^r_{i_1\cdots
i_{s-1}}(\vec{k})e_r\right]+\hspace{6cm}\\
+Q\left[w\alpha^p_{i_1\cdots
i_{s-1}}(\vec{k})e_p-iswb^r_{i_1\cdots
i_{s-1}}(\vec{k})e_r-2iw\sigma^r_{i_1\cdots
i_{s-1}}(\vec{k})e_r\right]=0\hspace{1cm}
\end{eqnarray*}
and its time part
\begin{eqnarray*}
P\left[w^2b^q_{i_1\cdots i_{s-1}}(\vec{k}) e_q + w^2b^r_{i_1\cdots
i_{s-1}}(\vec{k})e_r\right] +\hspace{7cm}\\
+Q\left[-iw^2 s^{-1}\alpha^p_{i_1\cdots
i_{s-1}}(\vec{k})e_p-w^2b^r_{i_1\cdots
i_{s-1}}(\vec{k})e_r-2w^2s^{-1}\sigma^r_{i_1\cdots
i_{s-1}}(\vec{k})e_r\right]=0\, ,
\end{eqnarray*}
which can be collected in the algebraic constraint
\begin{eqnarray*}
P\left[b^q_{i_1\cdots i_{s-1}}(\vec{k})e_q+b^r_{i_1\cdots
i_{s-1}}(\vec{k})e_r\right]=\hspace{7cm}
\\\hspace{3cm}=Q\left[ \frac{i}{s}\alpha^p_{i_1\cdots
i_{s-1}}(\vec{k}) +b^r_{i_1\cdots
i_{s-1}}(\vec{k})e_r+\frac{2}{s}\sigma^r_{i_1\cdots
i_{s-1}}(\vec{k})e_r\right]\, .
\end{eqnarray*}

\section*{Appendix D.}

In order to compute the symplectic structure we need
\begin{eqnarray*}
\gamma^q&=&\left\{\begin{array}{c}
\beta^q_{i_1\cdots i_s}(\vec{k})e^{-iwt}+\bar{\beta}^q_{i_1\cdots i_s}(-\vec{k})e^{iwt}\\ \\
       b^q_{i_1\cdots i_{s-1}}(\vec{k})e^{-iwt}+\bar{b}^q_{i_1\cdots
        i_{s-1}}(-\vec{k})e^{iwt}
\end{array}\right\}
\\
\gamma^p&=&\left\{\begin{array}{c}
\frac{i}{w}k_{[i_1}\left[\alpha^p_{i_2\cdots
i_s]}(\vec{k})e^{-iwt}
+\bar{\alpha}^p_{i_2\cdots i_s]}(-\vec{k})e^{iwt}\right]\\ \\
\frac{i}{w}k_{[i_1}\left[ a^p_{i_2\cdots
i_{s-1}]}(\vec{k})e^{-iwt}
    +\bar{a}^p_{i_2\cdots i_{s-1}]}(-\vec{k})e^{iwt}\right]
\end{array}\right\}
\\
\gamma^r&=&\left\{\begin{array}{l}
\beta^r_{i_1\cdots i_s}(\vec{k})e^{-iwt}+\bar{\beta}^r_{i_1\cdots i_s}(-\vec{k})e^{iwt}\\ \\
\frac{i}{w}k_{[i_1}\left[ a^r_{i_2\cdots
i_{s-1}]}(\vec{k})e^{-iwt}
    +\bar{a}^r_{i_2\cdots i_{s-1}]}(-\vec{k})e^{iwt}\right]+\\
\hspace{2cm}+b^r_{i_1\cdots i_{s-1}}(\vec{k})e^{-iwt}
    +\bar{b}^r_{i_1\cdots i_{s-1}}(-\vec{k})e^{iwt}
\end{array}\right\}
\\
\hspace{-2cm}d\delta\gamma^r_{{\scriptscriptstyle H\!D}}&\hspace{-.5cm}=&\hspace{-.5cm}
\left\{\begin{array}{l}
  \frac{i}{w}k_{[i_1}\left\{
    \left[\alpha^r_{i_2\cdots i_s]}(\vec{k})+wt\sigma^r_{i_2\cdots i_s]}
    (\vec{k})\right]e^{-iwt}
    +\hspace{2cm}\right.\\
\hspace{2cm}+\left.
 \left[\bar{\alpha}^r_{i_2\cdots i_s]}(-\vec{k})+wt\bar{\sigma}^r_{i_2\cdots i_s]}
 (-\vec{k})\right]e^{iwt}
 \right\}\\ \\
\frac{1}{s}\left[\left(\sigma^r_{i_1\cdots i_{s-1}}(\vec{k})
    -
    i\alpha^r_{i_1\cdots i_{s-1}}(\vec{k})-iwt\sigma^r_{i_1\cdots
    i_{s-1}}(\vec{k})\right)e^{-iwt}+\right.\hspace{1cm}\\
    \hspace{0.1cm}\left. +
    \left(\bar{\sigma}^r_{i_1\cdots i_{s-1}}(-\vec{k})
    +
    i\bar{\alpha}^r_{i_1\cdots i_{s-1}}(-\vec{k})+iwt\bar{\sigma}^r_{i_1\cdots
    i_{s-1}}(-\vec{k})\right)e^{iwt}\right]
\end{array}\right\}
\\
d\gamma^q&=&\left\{\begin{array}{c} ik_{[i_1} \left[
\beta^q_{i_2\cdots i_{s+1}]}(\vec{k})e^{-iwt}
+\bar{\beta}^q_{i_2\cdots
i_{s+1}]}(-\vec{k})e^{iwt} \right]\\ \\
ik_{[i_1}\left[-\frac{s}{s+1}b^q_{i_2\cdots
i_s]}(\vec{k})e^{-iwt}-\frac{s}{s+1}\bar{b}^q_{i_2\cdots
i_s]}(-\vec{k})e^{iwt}\right]-\\
\quad -\frac{iw}{s+1}\beta^q_{i_1\cdots
i_s}(\vec{k})e^{-iwt}+\frac{iw}{s+1}\bar{\beta}^q_{i_1\cdots
i_s}(-\vec{k})e^{iwt}
\end{array}\right\}
\\
d\gamma^r&=&\left\{\begin{array}{c} ik_{[i_1} \left[
\beta^r_{i_2\cdots i_{s+1}]}(\vec{k})e^{-iwt}
+\bar{\beta}^r_{i_2\cdots
i_{s+1}]}(-\vec{k})e^{iwt} \right]\\ \\
ik_{[i_1}\left[-\frac{s}{s+1}b^r_{i_2\cdots
i_s]}(\vec{k})e^{-iwt}-\frac{s}{s+1}\bar{b}^r_{i_2\cdots
i_s]}(-\vec{k})e^{iwt}\right]-\\
\quad-\frac{iw}{s+1}\beta^r_{i_1\cdots
i_s}(\vec{k})e^{-iwt}+\frac{iw}{s+1}\bar{\beta}^r_{i_1\cdots
i_s}(-\vec{k})e^{iwt}
\end{array}\right\}
\\ \delta \gamma^p&=&\left\{\begin{array}{c}
ik_{[i_1}\left[-isa^p_{i_1\cdots
i_{s-1}}(\vec{k})e^{-iwt}+is\bar{a}^p_{i_1\cdots
i_{s-1}}(-\vec{k})e^{iwt} \right]+\\\quad+w\alpha^p_{i_1\cdots
i_{s-1}}(\vec{k})e^{-iwt}+w\bar{\alpha}^p_{i_1\cdots
i_{s-1}}(-\vec{k})e^{iwt}\\ \\
-\frac{s}{s-1}wa^p_{i_1\cdots
i_{s-2}}(\vec{k})e^{-iwt}-\frac{s}{s-1}w\bar{a}^p_{i_1\cdots
i_{s-2}}(-\vec{k})e^{iwt}
\end{array}\right\}
\\
\delta \gamma^r&=&\left\{\begin{array}{c}
 ik_{[i_1}\left[-isa^r_{i_1\cdots
i_{s-1}}(\vec{k})e^{-iwt}+is\bar{a}^r_{i_1\cdots
i_{s-1}}(-\vec{k})e^{iwt} \right]-\\\quad-iswb^r_{i_1\cdots
i_{s-1}}(\vec{k})e^{-iwt}+isw\bar{b}^r_{i_1\cdots
i_{s-1}}(-\vec{k})e^{iwt}\\ \\
-\frac{s}{s-1}wa^r_{i_1\cdots
i_{s-2}}(\vec{k})e^{-iwt}-\frac{s}{s-1}w\bar{a}^r_{i_1\cdots
i_{s-2}}(-\vec{k})e^{iwt}
\end{array}\right\}
\\
\delta d\delta\gamma^r_{{\scriptscriptstyle
H\!D}}&=&\left\{\begin{array}{c} -2iw\sigma^r_{i_1\cdots
i_{s-1}}(\vec{k})e^{-iwt}+2iw\bar{\sigma}^r_{i_1\cdots
i_{s-1}}(-\vec{k})e^{iwt}\\0
\end{array}\right\}\, .
\end{eqnarray*}
The symplectic form associated to the action (\ref{003}) with ${\cal
M}=\real^4$ is
\begin{eqnarray}
\Omega_{\cal F}=\int_{\real^3}\left[ \dd A^\texttt{t}\w *P d\dd A+
(Q\delta \dd A)^\texttt{t}\w
*\dd A \right]\quad.\label{D10}
\end{eqnarray}
We must compute the restriction of (\ref{D10}) to the space of solutions
to the field equations $\cal S$. This gives
\begin{eqnarray*}
\left. \int_{\real^3}\left[ \dd A^\texttt{t}\w *P d\dd A+
(Q\delta \dd A)^\texttt{t}\w
*\dd A \right]\right|_{\cal S}=\hspace{4cm}\nonumber
\end{eqnarray*}
\vspace{-5mm}
\begin{eqnarray}
&\quad&\quad\quad=\int_{\real^3}(\dd\gamma^qe_q^{\texttt{t}}+
\dd\gamma^re_r^\texttt{t})\w
*P(d\dd\gamma^{q'}e_{q'}+d\dd\gamma^{r'}e_{r'})+\nonumber
\\
&\quad&\quad\quad+\int_{\real^3}d\dd\Lambda^qe_q^\texttt{t}
\w*P(d\dd\gamma^{q'}e_{q'}+d\dd\gamma^{r'}e_{r'})+\nonumber
\\
&\quad&\quad\quad+\int_{\real^3}(Q(\delta\dd
\gamma^pe_p+(\delta\dd\gamma^r+\delta d\delta
\dd\gamma^r_{{\scriptscriptstyle H\!D}})e_r))^\texttt{t}
\w*\delta\dd\Theta^pe_p+\nonumber
\\
&\quad&\quad\quad+\int_{\real^3} (Q(\delta \dd\gamma^pe_p+\delta
\dd\gamma^re_r))^\texttt{t}\w*(\dd\gamma^{p'}e_{p'}+\dd\gamma^{r'}e_{r'})+\label{D11}
\\
&\quad&\quad\quad+\int_{\real^3}(Q\delta d\delta
\dd\gamma^r_{{\scriptscriptstyle H\!D}}e_r)^\texttt{t}\w*(\dd\gamma^{p'}
e_{p'}+\dd\gamma^{r'}e_{r'})+
\nonumber\\
&\quad&\quad\quad+\int_{\real^3}(Q(\delta\dd
\gamma^pe_p+(\delta\dd\gamma^r+\delta d\delta
\dd\gamma^r_{{\scriptscriptstyle H\!D}})e_r))^\texttt{t}\w*d\delta
\dd\gamma^{r'}_{{\scriptscriptstyle H\!D}}e_{r'}+
\nonumber
\\
&\quad&\quad\quad+\int_{\real^3}d\delta\dd
\gamma^r_{{\scriptscriptstyle H\!D}}e_r^\texttt{t}
\w*P(d\dd\gamma^{q'}e_{q'}+d\dd\gamma^{r'}e_{r'})
\quad.\nonumber
\end{eqnarray}
To proceed further\footnote{Notice that $d$ and $\delta$ are four
dimensional operators but the integrals extend only to $\real^3$
so that we cannot integrate by parts.} we need the following
formula for a $s$-form $\ou$ and a $(s+1)$-form $\od$
\begin{eqnarray*}
\int_{\real^n}\ou\wedge*\od
&=&(s+1)!\int_{\real^n}\frac{d^n\vec{k}}{w^2}\beu_{i_1\cdots
i_s}(\vec{k},t)\bd _{i_1\cdots i_s}(-\vec{k},t)\\
&+&\frac{(s+1)!}{s}\int_{\real^n}d^n\vec{k}\alu_{i_1\cdots
i_{s-1}}(\vec{k},t)\ad_{i_1\cdots
i_{s-1}}(-\vec{k},t)\quad.\nonumber
\end{eqnarray*}
The first term in (\ref{D11}) can be easily computed just by
substituting the solutions to the field equations and taking into
account that integrals such as
\begin{eqnarray*}
\int_{\real^3} \frac{d^3\vec{k}}{w}\dd
\beta_{i_1\cdots i_s}^\texttt{t}(\vec{k})\w P\dd
\beta_{i_1\cdots i_s}(-\vec{k})=0\quad,
\end{eqnarray*}
due to the fact that $P$ is symmetric and $\dd \beta$ in a 1-form
in the solution space. This way we get
\begin{eqnarray*}
\int_{\real^3}\dd\gamma^\texttt{t}\w *Pd\dd\gamma=-2i
s!\int_{\real^3}\frac{d^3\vec{k}}{w}\dd\bar{\beta}^\texttt{t}_{i_1\cdots
i_s}(\vec{k})\w P\dd \beta_{i_1\cdots i_s}(\vec{k})\quad.
\end{eqnarray*}
The second term in (\ref{D11}) can be easily seen to be zero as a
consequence of the algebraic constraint and the fact that
$Qe_q=0$. The third term  is also zero because
\begin{eqnarray*}
Q\left[\delta \dd \gamma^pe_p+\left( \delta\dd \gamma^r+\delta
d\delta \dd\gamma^r_{{\scriptscriptstyle H\!D}}\right)e_r \right]
\end{eqnarray*}
can be written as the matrix $P$ acting on something and
$Pe_p=0$.
The computation of the sum of the fourth and fifth
term  is straightforward and gives
\begin{eqnarray*}
\int_{\real^3} (Q(\delta \dd\gamma^pe_p+\delta
\dd\gamma^re_r))^\texttt{t}\w*(\dd\gamma^{p'}e_{p'}+
\dd\gamma^{r'}e_{r'})+\hspace{3cm}\\
\hspace{3cm}+\int_{\real^3}(Q\delta d\delta
\dd\gamma^r_{{\scriptscriptstyle H\!D}}e_r)^\texttt{t}
\w*(\dd\gamma^{p'}e_{p'}+\dd\gamma^{r'}e_{r'})=\\
 =\frac{2is s!}{s-1}\int_{\real^3}\frac{d^3\vec{k}}{ w}\dd
\bar{a}^\texttt{t}_{i_1\cdots i_{s-2}}(\vec{k})\w Q\dd
a_{i_1\cdots i_{s-2}}(\vec{k}) +\hspace{3cm}
\\\hspace{3cm}+2iss!\int_{\real^3}\frac{d^3\vec{k}}{w}\dd
\bar{b}^r_{i_1\cdots i_{s-1}}e_r^\texttt{t}(\vec{k})\w P\dd
b_{i_1\cdots i_{s-1}}(\vec{k})\quad.
\end{eqnarray*}
Finally the last two terms require a somewhat lengthy computation
that is simplified by the following remarks:

First of all, the terms proportional to $e^{2iwt}$ and $e^{-2iwt}$
are both zero. Second \vspace*{-5mm}$$
\int_{\real^3}\frac{d^3\vec{k}}{ w}\left\{\dd \sigma_{i_1\cdots
i_{s-1}}^r(\vec{k}) e_r^\texttt{t}\w P\left[\dd \bar{b}_{i_1\cdots
i_{s-1}}^q(\vec{k})e_q+ \dd\bar{b}_{i_1\cdots
i_{s-1}}^r(\vec{k})e_r\right]+\right.$$ $$\hspace{2cm}+ \left.\dd
\bar{\sigma}_{i_1\cdots i_{s-1}}^r(\vec{k}) e_r^\texttt{t}\w
P\left[\dd b_{i_1\cdots i_{s-1}}^q(\vec{k})e_q+ \dd b_{i_1\cdots
i_{s-1}}^r(\vec{k})e_r\right]\right\}=0 $$ as can be seen by
using the algebraic constraint to write
\begin{eqnarray*}
P\left[\dd b^q_{i_1\cdots i_{s-1}}(\vec{k})e_q+\dd b^r_{i_1\cdots
i_{s-1}}(\vec{k})e_r\right]=\hspace{6.5cm}
\\\hspace{2.5cm}=Q\left[ \frac{i}{s}\dd\alpha^p_{i_1\cdots
i_{s-1}}(\vec{k}) +\dd b^r_{i_1\cdots
i_{s-1}}(\vec{k})e_r+\frac{2}{s}\dd\sigma^r_{i_1\cdots
i_{s-1}}(\vec{k})e_r\right]\quad,
\end{eqnarray*}
and the fact that $\dd A{\wedge \kern-.57em \wedge}\dd B=-\dd
B{\wedge \kern-.57em \wedge}\dd A$ for 1-forms.

So we obtain
\begin{eqnarray*}
& &\int_{\real^3}d\delta\dd\gamma^r_{{\scriptscriptstyle
H\!D}}e_r^\texttt{t}
\w*P(d\dd\gamma^{q'}e_{q'}+d\dd\gamma^{r'}e_{r'})+
\\
& &\hspace{2.4cm}+\int_{\real^3}(Q(\delta\dd
\gamma^pe_p+(\delta\dd\gamma^r+\delta d\delta
\dd\gamma^r_{{\scriptscriptstyle H\!D}}e_r))^\texttt{t}
\w*d\delta\dd\gamma^{r'}_{{\scriptscriptstyle H\!D}}e_{r'}=
\\
& =& is!\int_{\real^3}\frac{d^3 \vec{k}}{w}
\,\left[\dd\bar{\sigma}^r_{i_1\cdots
i_{s-1}}(\vec{k})e_r^\texttt{t}\w P\dd b_{i_1\cdots
i_{s-1}}(\vec{k})-\dd \sigma^r_{i_1\cdots
i_{s-1}}(\vec{k})e_r^\texttt{t}\w P\dd \bar{b}_{i_1\cdots
i_{s-1}}(\vec{k})\right]
\\
& -&2s! \int_{\real^3}\frac{d^3 \vec{k}}{w}\,\left[
\dd\bar{\alpha}_{i_1\cdots i_{s-1}}^r(\vec{k})e_r^\texttt{t}\w
P\dd b_{i_1\cdots i_{s-1}}(\vec{k})+\dd \alpha_{i_1\cdots
i_{s-1}}^r(\vec{k})e_r^\texttt{t}\w P\dd\bar{b}_{i_1\cdots
i_{s-1}}(\vec{k}) \right].
\end{eqnarray*}
Adding up all these contributions we finally get (\ref{030}).

\section*{Appendix E.}

If a diagonal element in a quadratic form ${\cal Q}$ is zero, and
the row and column where this element is are not identically zero,
then $\cal Q$ cannot be either definite or semidefinite. The
proof is very simple; let us write $${\cal Q}=\left[
\begin{array}{cc}
    0& v^\texttt{t}\\v&q
\end{array}
\right]$$
then
\begin{equation}
X^\texttt{t} {\cal Q} X=[x^\texttt{t} \,\,\,y^\texttt{t}]\left[
\begin{array}{cc}
    0& v^\texttt{t}\\v&q
\end{array}
\right]\left[
\begin{array}{c}
    x\\y
\end{array}
\right]=2y^\texttt{t} v x+y^\texttt{t} q y. \label{E}
\end{equation}
Let us fix $y$ such that $y^\texttt{t} v\neq 0$. We have then that
(\ref{E}) can take both positive and negative values depending on
the choice of $x$.

\end{document}